\documentclass[11pt]{article}
\usepackage{hyperref,graphicx,amsthm,amsfonts,amsmath,authblk} 
\usepackage[sort&compress,comma,numbers]{natbib}
\usepackage[top=2cm, bottom=2cm, outer=2cm, inner=2cm, heightrounded]{geometry}
\usepackage{color}
\newtheorem{theorem}{Theorem}[section]
\newtheorem{lemma}{Lemma}[subsection]
\newcommand{\makeSymbol}[1]{\mathord{\vcenter{\hbox{#1}}}}
\newcommand{\z}{\hat{\mathcal{Z}}}
\newcommand{\f}{\mathcal{F}}
\newcommand{\ada}{\hat{a}^\dagger}
\newcommand{\haa}{\hat{a}}
\newcommand{\n}{\hat{N}}

\newcommand{\h}{\hat{H}_v}
\newcommand{\hH}{\hat{H}}

\newcommand{\dualvec}[1]{\Big(#1\Big|}
\newcommand{\dep}{deparametrized }
    
\begin{document}
\title{Towards the self-adjointness of a Hamiltonian operator in loop quantum gravity}
\author[1,2]{Cong Zhang\thanks{zhang.cong@mail.bnu.edu.cn}}
\author[1,2]{Jerzy Lewandowski\thanks{jerzy.lewandowski@fuw.edu.pl}}
\author[1]{Yongge Ma\thanks{mayg@bnu.edu.cn}}

\affil[1]{ Department of Physics, Beijing Normal University, Beijing 100875, China}
\affil[2]{Faculty of Physics, University of Warsaw, Pasteura 5, 02-093 Warsaw, Poland}

\date{}
\maketitle 
\begin{abstract}
Although the physical Hamiltonian operator can be constructed in the \dep model of loop quantum gravity coupled to a scalar field, its property is still unknown. This open issue is attacked in this paper by considering an operator $\h$ representing the square of the physical Hamiltonian operator acting nontrivially on a two-valent vertex of spin networks. The Hilbert space $\mathcal{H}_v$ preserved by the graphing changing operator $\h$ is consist of spin networks with a single two-valent non-degenerate vertex. The matrix element of $\h$ are explicitly worked out in a suitable basis. It turns out that the operator $\h$ is essentially self-adjoint, which implies a well-defined physical Hamiltonian operator in $\mathcal{H}_v$ for the \dep model.\\
{ }
\end{abstract}

\section{Introduction} 
Loop Quantum Gravity (LQG)  is a background independent framework designed for quantization of generally relativistic theories of gravity coupled to other fields \cite{ashtekar2004back,han2007fundamental,rovelli2005quantum,thiemann2007modern}. In this paper we consider the canonical approach as opposed to  the covariant Spin-Foam models.
The starting point of the canonical LQG is  the standard,  torsion  free Einstein's gravity in the Palatini-Holst \cite{holst1996barbero,ashtekar2004back} formulation coupled to the fields of the Standard Model of fundamental interactions. The  peculiar step, though,   is a reformulation of the  canonical classical theory in terms of the Ashtekar-Barbero variables: an SU(2) connection $A$ and the canonically conjugate  $3$-frame-density variable $E$
\cite{ashtekar1991lectures,barbero1995real}. The second  peculiar step is the introduction of the parallel transport (holonomy) of $A$ (along all the curves) and flux of $E$ (along all the $2$-surfaces) as basic variables subject to the canonical quantization \cite{rovelli1988knot}. The quantization provided several important break-throughs.  A  Hilbert space was defined that caries a unitary action of the spatial diffeomorphisms and quantum representation of the holonomy-flux variables \cite{ashtekar1994representation,lewandowski1994topological,ASHTEKAR1995191}.  A family of operators representing   geometric observables  ($2$-surface area, $3$-region volume, inverse metric tensor) were regularized without need to subtract infinities and their spectra turned out to be discrete \cite{rovelli1995discreteness,ASHTEKAR1995191,ashtekar1997quantum,ashtekar1997quantumII,ma2000qhat,yang2016new}.  The representation is unique upon the diffeomorphism invariance and the existence of an invariant cyclic state \cite{lewandowski2005uniqueness}. The Gauss and vector constraint were solved exactly and the "half-physical" space of solutions was endowed with a natural Hilbert product \cite{ashtekar1995quantization}.  Thereon the quantum scalar   constraint map was regularized   again without emergence of any infinities that would have to be abandoned \cite{thiemann1998quantum,thiemann1998quantumII, ashtekar2004back}. The Master Constraint operator that is supposed to capture  all the constraints of the vacuum theory was defined \cite{thiemann2006phoenix,han2006master}.  Finally, a symmetric (in the sense of the Hilbert product) quantum representation for the gravitational part of the scalar constraints was introduced in a suitably defined  vertex Hilbert space \cite{lewandowski2014symmetric,assanioussi2015new,yang2015new}.  Also matter fields were coupled to LQG and quantizations  consistent with the new framework were found. In particular,  the Brown-Kuchar model of gravity coupled to dust as well as the Rovelli-Smolin model of gravity coupled to massless Klein-Gordon field were quantized completely  \cite{rovelli1994physical,brown1995dust,giesel2010algebraic,domagala2010gravity,lewandowski2011dynamics}.  The mechanisms of swallowing gauge dependent degrees of freedom, constructing and evolving  Dirac observables in a relational manner were discussed, understood better and reformulated \cite{rovelli1991observable,dittrich2006partial,thiemann2006reduced,dapor2013relational}.  

The curvature of the Ashtekar-Barbero connection present in a quantum 
gravitational Hamiltonian acts on quantum states of the canonical LQG by 
attaching loops. A specific  way the quantum curvature  does it  is not 
determined uniquely, it can be defined in infinitely many different and 
inequivalent ways \citep{ashtekar2004back}.  That causes a considerable  ambiguity in 
defining a quantum Hamiltonian operator.  A rough  classification  
divides the set of all the Hamiltonian operators of LQG into the 
following two categories: $(i)$ graph preserving, $(ii)$  graph non-preserving.
The graph preserving action is natural from the lattice discretization 
point of view.  It makes the action of operators reducible to subspaces 
corresponding to the graphs. For every graph,  analytic properties of 
operators are much easier to study.  There is also a general argument, 
that if gravity is deparametrized by a coupled  dust, then the graph 
preserving action is the only diffeomorphism invariant option \cite{giesel2010algebraic}. The 
discretization requires taking a continuum limit. That can be achieved 
by  a suitable renormalization scheme  \cite{lang2017hamiltonian,lang2017hamiltonianII}.  Remarkably, the 
renormalization  attempts also to resolve the remaining quantization 
ambiguities of the Hamiltonian operator.  A continuum field theory 
approach rather than the discretization,   leads directly to the second 
category, that is to the
graph-changing action.  This is the option our current  paper concerns.  
  Several proposals of graph changing quantum Hamiltonian operator were 
considered in the literature \cite{ashtekar2004back,giesel2010algebraic,thiemann2007modern,alesci2010regularization,alesci2015hamiltonian,assanioussi2015new}.   A requirement 
that the Hermitian adjoint operator to an operator adding a loop is a 
well defined operator imposes conditions on admissible  ways of 
attaching loops \cite{lewandowski2014symmetric,yang2015new}.     Still, however, nothing was known thus far 
about the self-adjointness. In quantum mechanics the self-adjointness of quantum observables
corresponds to the  reality of corresponding  classical observables, hence it has a clear physical meaning. The self-adjoitness 
of an operator 
is equivalent to the spectral decomposition and reality of the spectrum that allows to define a given operator by indicating eigenstates and  corresponding eigenvalues. In particular, this is the spectral decomposition of quantum constraints that  ensures exact definition of their solutions  and endows their space with a natural Hilbert product.  The self-adjoitness of effective Hamiltonian operators ensures existence of a unitary time evolution of quantum states.  The relevance of that property of the quantum Hamiltonian is illustrated  in the models of loop quantum cosmology \cite{ashtekar2011loop,ashtekar2003mathematical}.  The big breakthrough of that theory coming after  a genuine  self-adjoint quantum  Hamiltonian operator was introduced in \cite{ashtekar2006quantumnature,kaminski2008flat}. 

In the current paper, for the first time, we address the issue of the self-adjointness of the graph changing Hamiltonian in the full theory with the local degrees of freedom. The model we choose to study is LQG coupled to the massless Klein-Gordon field.  
This is the full set of degrees of freedom version of the Ashtekar-Pawłowski-Singh symmetry reduced homogeneous-isotropic model of universe.  That is also one of the two known  remarkable cases  in which the Dirac program of quantum gravity can be completed \cite{domagala2010gravity,lewandowski2011dynamics} (the second case after the Brown-Kuchar model). Indeed, all the quantum constraints of the canonical General Relativity were solved completely and a 
general solution was written down explicitly, assuming the existence of certain operators.  The physical Hilbert space of the solutions was defined.  The general formula for  a Dirac observable that commutes with all the constraints  was derived.  The resulting algebra of the Dirac observables  was shown to admit an action of the $1$-dimensional group of automorphisms  that classically corresponds to the transformations of adding  a constant to the scalar field.  The generator of those automorphisms  was promoted to the physical quantum Hamiltonian operator $\hat{H}_{\rm phys}$ of the system. 
 An exact derivation of that operator in LQG has become possible with the introduction of the vertex Hilbert spaces \cite{lewandowski2014symmetric,yang2015new}.  The advantage of that Hilbert space necessary for our purpose is that it admits quantum operators of the gravitational scalar constraint 
 $\hat{C}^{\rm gr}(x)$  smeared against any suitable test function $N$ on the spatial manifold $\Sigma$ as
$$\hat{C}^{\rm gr}(N)\ =\ \int_\Sigma d^3xN(x) \hat{C}^{\rm gr}(x).$$ 
An existence of the Hermitian adjoint operator  $\left({\hat{C}_{\rm gr}}(N)\right)^\dagger$ is a condition on the quantization. It is satisfied by two alternative
proposals. The first one changes the valency of the vertices, but remains the spins on  the old edges invariant \cite{lewandowski2014symmetric}. The second one preserves the valency of the vertices and changes the spins of the edges in a way controlled to the effect that 
no spin can be reduced to zero \cite{yang2015new}.  In the current paper we apply the latter proposal. We  combine it  with the new idea of quantization of the gravitational scalar constraint \cite{Domagala2015kse, alesci2015hamiltonian}.  

Due to that choice a physical Hamiltonian operator can be constructed without using the quantum volume operator \cite{ashtekar1997quantum}.  In the consequence, this physical Hamiltonian even does not annihilate the spin network states with two-valent vertices since it does not contain the volume operator. This helps us to find an example  of a simple subspace of the full Hilbert space that is
preserved by the action of the quantum Hamiltonian and analyse the restricted operator. The subspace is constructed by introducing a graph with a single two-valent vertex. A quantum state defined by a 2-valent vertex  classically corresponds to a degenerate Ashtekar frame  such that one of the densitized vectors  is zero. The phase space of the classical Ashtekar theory contains points characterized by the lower than 3 rank of the frame. Mathematically, they are regular and  make perfect sense \cite{ma1998degenerate,ma1999causal}.  In particular, the rank 2 case is exactly soluble for the vacuum theory \cite{lewandowski19972}. The result is a generalized spacetime foliated by disjoined  2+1 
dimensional surfaces. Quantum states defined by graphs containing only 
2-valent vertices are quantization of that degenerate sector of 
Ashtekar's theory.

 The paper is organized as follows.   In Sec. \ref{se:one}, we remind the classical and quantum theories of \dep model of gravity coupled to massless Klein-Gordon scalar field. Some necessary  notions, like kinematical Hilbert space, the vertex Hilbert space and the physical Hilbert space, are included. The key part of this section is the explicit definition of the physical Hamiltonian operator on the vertex Hilbert space. In Sec. \ref{se:two}, we apply the general theory to a simple case, where the Hilbert space is generated  from a single bivalent non-degenerate vertex and preserved by the physical Hamiltonian operator. The action of the operator $\hat{H}_v$ representing the square of the physical Hamiltonian  on the Hilbert space are derived.
 In Sec. \ref{se:three}, we study  the operator $\hat{H}_v$ on the simple subspace and prove that the restricted operator  is well defined and  self-adjoint.  In Sec. \ref{se:four}, we discuss the issue of eigenvalue problem for $\hat{H}_v$. Conclusions and outlooks are presented in Sec. \ref{se:five}.

\section{A general work on \dep model}\label{se:one}
\subsection{The classical theory}
 Considering gravity minimally coupled to a massless Klein-Gordon field in the ADM formalism with Ashtekar-Barbero variables, we have a totally constrained system with the standard canonical variables $(A^i_a(x),E^a_i(x))$ for gravity and $(\phi(x),\pi(x))$ for scalar field defined at every point $x$
 of an underlying $3$-dimensional manifold $\Sigma$. The diffeomorphism and scalar constraints are respectively
\begin{equation}\label{eq:vectorconstraint}
C_a(x)=C_a^{\rm gr}(x)+\pi(x)\phi_{,a}(x)=0,
\end{equation}
\begin{equation}\label{eq:scalarconstraint}
C(x)=C^{\rm gr}(x)+\frac{1}{2}\frac{\pi(x)^2}{\sqrt{|\det E(x)}|}+\frac{1}{2}q^{ab}(x)\phi_{,a}(x)\phi_{,b}(x)\sqrt{|\det E(x)|}=0,
\end{equation}
where $C^{\rm gr}_a$ and $C^{\rm gr}$ are the vacuum gravity constraints and $q^{ab}=\frac{E^a_iE^b_i}{|\det E|}$. 

The \dep procedure starts with assuming that the constraints \eqref{eq:vectorconstraint} are satisfied. By replacing $\phi_{,a}$  by $-C_a^{\rm gr}/\pi$, the constraints \eqref{eq:scalarconstraint} are rewritten as
\begin{equation}
\pi^2=\sqrt{|\det E|}\left(-C^{\rm gr}\pm \sqrt{(C^{\rm gr})^2-q^{ab}C^{\rm gr}_aC^{\rm gr}_b}\right).
\end{equation}
 The sign ambiguity is solved depending on a quarter of the phase space. We choose the one that contains the homogeneous cosmological solutions \cite{alesci2015hamiltonian}.  
 In that part of the phase space, the scalar constraint $C(x)$ can be replaced by,  
 \begin{equation}
C'(x)=\pi(x)\pm \sqrt{h(x)},
\end{equation}
where
\begin{equation}\label{eq:classicalhamil}
h=\sqrt{|\det E|}\left(-C^{\rm gr}\pm \sqrt{(C^{\rm gr})^2-q^{ab}C^{\rm gr}_aC^{\rm gr}_b}\right).
\end{equation}
\subsection{The structure of the quantum theory}
For the \dep  theory, the Dirac quantization scheme can be implemented and performed to the end  \cite{domagala2010gravity,lewandowski2011dynamics}.  The result is a physical Hilbert space of solutions to the constraints, together with algebra of quantum Dirac observables endowed with one dimensional group of automorphisms generated by a quantum Hamiltonian operator. This resulted structure is equivalent to  the following
model that is expressed in a derivable way by elements of the framework of LQG:
\begin{itemize}
\item The physical Hilbert space $\mathcal{H}$ is the space of the quantum states of the vacuum (matter free) gravity
in the Ashtekar-Barbero connection-frame variables that satisfy the vacuum quantum vector constraint and the vacuum quantum Gauss constraint.  
In other words, in the connection representation,  the states are constructed from functions $A\mapsto \Psi(A)$ invariant with respect to the 
diffeomorphism gauge transformations
$$ A' = f^*A, \ \ \ \ \ \ \ \ \ \ \ \ \ f\in {\rm Diff}(\Sigma)$$
and to the Yang-Mills gauge transformations 
\begin{equation} \label{YM}
 A' = g^{-1}Ag + g^{-1} dg, \ \ \ \ \ \ \ \ \ \ \ \ \ g\in C(\Sigma,G).
  \end{equation}
They are not assumed to satisfy the vacuum scalar constraint, though. That Hilbert space is available in the LQG framework.  

\item The Dirac observables are represented by the set of operators $\{\hat{\mathcal{O}}\}$  in $\mathcal{H}$. When the scalar field transforms as $\phi\mapsto\phi+t$ with a constant $t$, the observables transform as
\begin{equation}
\hat{\mathcal{O}}\mapsto e^{i\hH t}\hat{\mathcal{O}} e^{-i\hH t}
\end{equation}
Therefore the quantum dynamics in the Schr\"odinger picture  is given by
\begin{equation}
 i\frac{d}{dt}\Psi=\hH\Psi,
 \end{equation} 
 $\hH$ is called the quantum Hamiltonian.

 \item The quantum Hamiltonian 
 \begin{equation}
 \hH=\int d^3x \widehat{\sqrt{-2\sqrt{|\det E(x)|}C^{\rm gr}(x)}}
 \end{equation}
 is a quantum operator corresponding to the classical physical Hamiltonian $$H=\int d^3x \sqrt{-2\sqrt{|\det E(x)|}C^{\rm gr}(x)}.$$ 
\end{itemize} 
 
The classical Hamiltonian $H$  is manifestly spatial diffeomorphism invariant, the same is expected  about a quantum Hamiltonian operator 
$\hH$. There seems to be a perfect compatibility between the diffeomorphism invariance of the quantum Hamiltonian operator and the diffeomorphism invariance of the quantum states, elements of the Hilbert space ${\cal H}$.  However,   the integrant $\sqrt{-2\sqrt{|\det E(x)|}C^{\rm gr}(x)}$ involves the square root of an expression  assigned to each point $x$. In order to quantize it, we should know the operator corresponding to the expression of $-2\sqrt{|\det E(x)|}C^{\rm gr}(x)$  under the square root at first. However, on one hand $-2\sqrt{|\det E(x)|}C^{\rm gr}(x)$ itself is not diffeomorphism invariant, which leads to the fact that the corresponding operator can not be well defined within the diffeomorphism invariant Hilbert space. On the other hand, group averaging with respect to diffeomorphism transformation is necessary because the operator corresponding to $-2\sqrt{|\det E(x)|}C^{\rm gr}(x)$ is defined by taking some limit of holonomies along a sequence of closed loops nearby the vertices of spin networks. The convergence of such limits requires  partial diffeomorphism invariance nearby each vertex. Therefore, the kinematical Hilbert space is not a suitable choice either. The idea to solve the contradiction is to introduce the vertex Hilbert space $\mathcal{H}_{\rm vtx}$, of partially diffeomorphism invariant  states  in which an operator $ \widehat{-2\sqrt{|\det E(x)|}C^{\rm gr}(x)}$ is well defined for each $x\in\Sigma$  
 \cite{lewandowski2014symmetric,yang2015new,alesci2015hamiltonian} . We can finally pass to  the operators in $\mathcal{H}$ from those in $\mathcal{H}_{\rm vtx}$ by the dual action naturally since $\mathcal{H}$ is a dual space of $\mathcal{H}_{\rm vtx}$ as shown in the following context.
   
\subsection{The Hilbert spaces  ${\cal H}_{\rm kin}$,  $\mathcal{H}_{\rm vtx}$ and ${\cal H}$} 
The kinematical Hilbert space ${\cal H}_{\rm kin}$ of the vacuum LQG  consists of functions 
\begin{equation}\label{cyl}
\Psi_\gamma(A)\ =\ \psi_\gamma(h_{e_1}(A), ... , h_{e_n}(A)),
\end{equation}
where $e_1,...,e_n$ are the edges of a graph embedded in $\Sigma$, and  $h_{e}(A)\in SU(2)$ is the parallel transport along 
a path $e$ in $\Sigma$ with respect to a given connection $1$-form $A$,
$$ h_e(A)\ =\mathcal{P} \exp(- \int_e A). $$
  In the LQG framework those functions of the variable $A$ are called cylindrical functions.
It may be also used to  define a multiplication operator, given a representation $D^{(l)}$ of SU(2), 
$$ (D^{(l)}{}^m{}_n (h_p)) \Psi(A)\ =\  D^{(l)}{}^m{}_n (h_p(A))\Psi(A),$$
where $m,n$ label an entry of the matrix. 

The kinematical space can be decomposed  into the orthogonal sum 
\begin{equation}\label{decompkin} 
{\cal H}_{\rm kin}\ =\ \overline{\bigoplus_\gamma {\cal H}_\gamma},
\end{equation}
where $\gamma$ runs through the set of embedded graphs in $\Sigma$ (un-oriented, and  without removable vertices).

We also use a basis $\tau_1,\tau_2,\tau_3 \in{\rm su}(2)$ such that
$$ [\tau_i,\tau_j]\ =\ \epsilon_{ijk}\tau_k . $$ 
Another operator we will apply in the current paper is defined in ${\cal H_\gamma}$. Given a graph $\gamma$, a vertex $v$, 
and an edge $e_0$  at $v$, and  $\tau_i$, it acts  on  function  $\Psi_\gamma$ of \eqref{cyl} as follows,
\begin{equation}\label{J}
(J_{v,e_0}^i\Psi_\gamma)(A)=\left\{
\begin{matrix}
i \left.\frac{d}{dt}\right|_{t=0}\psi_\gamma(\cdots,h_e(A),e^{ -t\tau_i}h_{e_0}(A),g_{e'}(A),\cdots),& v=t(e)\\
i \left.\frac{d}{dt}\right|_{t=0}\psi_\gamma(\cdots,h_e(A),h_{e_0}(A)e^{t\tau_i},h_{e'}(A),\cdots),& v=s(e).
\end{matrix}
\right.
\end{equation}
 see \cite{ashtekar2004back} for more details.

In this paper we restrict to   functions  invariant with respect to  the Yang-Mills gauge transformations (\ref{YM}).  
An orthonormal basis can be constructed from the spin-network states. 
Given a graph $\gamma$ in $\Sigma$, we denote by $V(\gamma)$ the set of the vertices and $E(\gamma)$ the set of the edges.  
A vertex $v\in V(\gamma)$ is called degenerate if all of the edges $e$ at $v$ are tangent to each other\footnote{This definition is from our quantization of the physical Hamiltonian in the following.}.  We denote the set of all non-degenerate vertices by $V_{\rm nd}(\gamma)$, the diffeomorphisms acting trivially on
$V_{\rm nd}(\gamma)$ by ${\rm Diff}^\omega_{\rm nd}(\gamma)$ with $\omega$ representing that the diffeomorphism  is semi-analytic \cite{lewandowski2005uniqueness},  and the elements of ${\rm Diff}^\omega_{\rm nd}(\gamma)$   preserving every edge of $\gamma$  by ${\rm TDiff}(\gamma)$. For any cylindrical function $\Psi_\gamma\in{\cal H}_\gamma$, the map 
$\eta$ is defined as
\begin{equation}\label{eta}
 \eta:\Psi_\gamma\mapsto \sum_{f\in \rm Diff^\omega_{nd}(\gamma)/TDiff(\gamma)} \langle \hat{U}_f\cdot \Psi_\gamma|,
 \end{equation} 
 where $\hat{U}_f$ denotes the unitary operator corresponding to the diffeomorphism transformation $f$ on $\Sigma$ \cite{ashtekar2004back}.
$\eta$ maps all elements in  ${\cal H}_\gamma$ into the algebraic dual $\left(\bigoplus_\gamma {\cal H}_\gamma\right)'$. The inner product in $\eta\left(\bigoplus_\gamma {\cal H}_\gamma\right)$ is defined naturally by
$$\Big( \eta(\Psi_\gamma),\eta(\phi_{\gamma'})\Big)=\eta(\Psi_\gamma)[\phi_{\gamma'}].$$ 
The resulting space $\eta\left(\bigoplus_\gamma {\cal H}_\gamma\right)$ admits the natural orthogonal
decomposition 
$$ \eta\left(\bigoplus_\gamma {\cal H}_\gamma\right)\ =\ \bigoplus_{[\gamma]} \eta\left({\cal H}_\gamma\right) ,$$
where $[\gamma]$ stands for the set of all the graphs $\gamma'$ such that
$$ \eta\left({\cal H}_{\gamma'}\right)\ =\  \eta\left({\cal H}_\gamma\right). $$ 
The vertex Hilbert space $\mathcal{H}_{\rm vtx}$ is the completion of $\eta\left(\bigoplus_\gamma {\cal H}_\gamma\right)$ under this inner product. One can conclude easily that for every graph $\gamma$, every element $\eta(\Psi_\gamma)\in \eta\left( {\cal H}_\gamma\right)$ is a partial solution to the quantum diffeomorphism constraint invariant with respect to all the diffeomorphisms contained 
in   ${\rm Diff}^\omega_{\rm nd}(\gamma)$. It can be turned into a full solution of the quantum diffeomorphism constraint by a similar averaging with respect to the remaining quotient space
$\rm Diff_\omega(\Sigma)/Diff^\omega_{\rm nd}(\gamma)$, which equals to the set of embeddings of $V(\gamma)$ in $\Sigma$. In this way 
the Hilbert space ${\cal H}$  mentioned above is defined as a dual space of $ \mathcal{H}_{\rm vtx}$.   Passing a diffeomorphism invariant operator from $\mathcal{H}_{\rm vtx}$ to 
$\mathcal{H}$ is naturally realized by the dual action. Therefore, without losing the generality, we will study  the quantum Hamiltonian operator 
$\hH$  in the Hilbert space $\mathcal{H}_{\rm vtx}$.  

Given an operator $\hat{\mathcal{O}}$ on $\mathcal{H}_{\rm kin}$, the corresponding operator 
$\hat{\mathcal{O}}'$ on $\mathcal{H}_{\rm vtx}$ is defined by the duality 
\begin{equation}\label{eq:dualaction}
\left(\hat{\mathcal{O}}'\eta(\Psi_\gamma)\right)[f]:=\eta(\Psi_\gamma)[\hat{\mathcal{O}}f],\ \ \ \ \ \ \ \ \forall f\in\bigoplus_\gamma {\cal H}_\gamma.
\end{equation}
 
\subsection{The physical quantum Hamiltonian operator}\label{se:phyH} 
In the current paper we combine the general regularization scheme for 
the operator $ \widehat{\sqrt{-2\sqrt{|\det E(x)|}C^{\rm gr}(x)}}$  introduced in \cite{alesci2015hamiltonian} with the vertex valency preserving  proposal for the loop assignment in \cite{yang2015new}. Thus the resulting 
operator is defined in the version of \cite{yang2015new} in the vertex Hilbert space introduced above.   
 
According to the framework, the integrant in the formula for the physical Hamiltonian operator is defined on the dual space 
$\left(\bigoplus_\gamma {\cal H}_\gamma\right)'$ and takes the following form,
\begin{equation}\label{theoperator}
\widehat{\sqrt{-2\sqrt{|\det E(x)|}C^{\rm gr}(x)}}\ =\ \sum_{v\in \Sigma}\delta(v,x)\sqrt{\h}
\end{equation}
where $\delta(v,x)$ is the Dirac distribution.
The sum seems to be awfully infinite. However, for every subspace $\eta\left({\cal H}_\gamma\right)$, the only non-zero
terms correspond to the vertices of the underlying graph $\gamma$ of a cylindrical function.
For every vertex $v\in V_{\rm nd}(\gamma)$  the operator $\h$ is defined first  as an operator ${}^{\rm kin}\h$ 
in the kinematical Hilbert subspace ${\cal H}_\gamma$, next pulled back by $\eta$ to a well-defined operator in 
$\eta\left({\cal H}_\gamma\right)$, and finally symmetrized. The issue is self-adjointness.  
 ${}^{\rm kin}\h$  takes the form  of the sum with respect to pairs of edges of 
$\gamma$ that meet at $v$, 
\begin{equation}\label{eq:physicalH}
{}^{\rm kin}\h = \sum_{e,e'\text{ at } v}\epsilon(e,e'){}^{\rm kin}\hH_{v,ee'}
\end{equation}
where $\epsilon(e,e')$ equals to $0$ if $e$ and $e'$ are tangent  at $v$ or $1$ otherwise.       
The operator consists of two parts,
$${}^{\rm kin}\hH_{v,ee'}\ =\ (1+\beta^2){}^{\rm kin}\hH^L_{v,ee'}+ {}^{\rm kin}\hH^E_{v,ee'},
$$ 
where  the operators ${}^{\rm kin}\hH^E_{v,ee'}$ and ${}^{\rm kin}\hH^L_{v,ee'}$ act as follows:
\begin{itemize}
\item ${}^{\rm kin}\hH^E_{v,ee'}$ creates a pair of vertices $v_L$ and $v_{R}$ that split $e$, and $e'$, respectively, 
and attaches a new edge $\ell$ connecting the new vertices,
\begin{equation} 
{}^{\rm kin}\hH^E_{v,ee'}\makeSymbol{\raisebox{0.\height}{\includegraphics[width=0.1\textwidth]{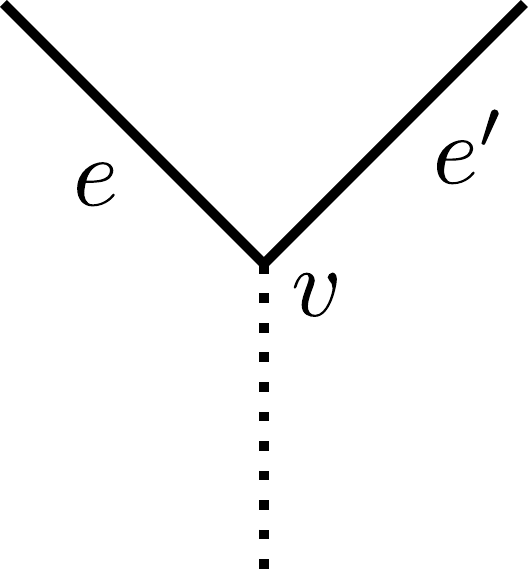}}}\to\ \makeSymbol{\raisebox{0.\height}{\includegraphics[width=0.1\textwidth]{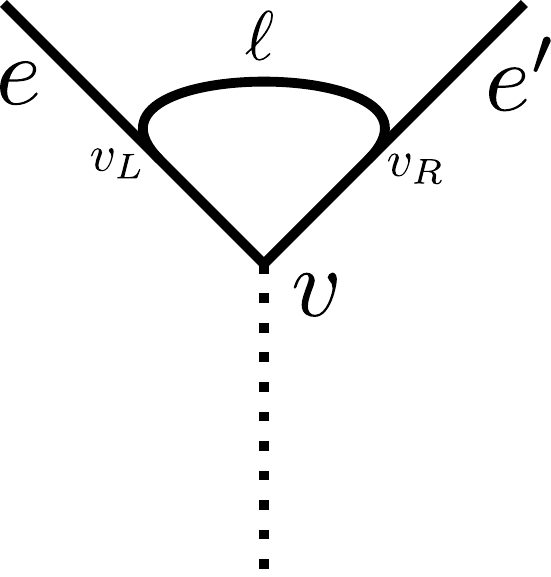}}}.
\end{equation}
A new element in this definition is that the new edge $\ell$ is tangent to both, $e$ at $v_L$, and $e'$ at $v_R$.
More specifically 
\begin{equation}\label{eq:eucleadianH}
{}^{\rm kin} \hH^E_{v,ee'}\ =\ \kappa_1\epsilon_{ijk}(h^i_{\alpha_{ee'}})^{(l)}\hat{J}^j_{v,e}\hat{J}^k_{v,e'},
\end{equation}    
where $\alpha_{ee'}$ is the loop passing through the vertices $v,v_L,v_R,v$ in the given order and along the segments of $e$ and $e'$, and along $\ell$,    and
\begin{align} 
(h^i_{\alpha_{ee'}})^{(l)} \ &:=\ -\frac{3}{l(l+1)(2l+l)}{\rm Tr}^{(l)}(h_{\alpha_{ee'}}\tau^i),\\
{\rm Tr}^{(l)}(h_{\alpha_{ee'}}\tau^i)\ &:=\ {\rm Tr} \left(D^{(l)}(h_{\alpha_{ee'}}) D'^{(l)}(\tau^i)\right).
\end{align}
The factor $\kappa_1$ is arbitrary, representing a residual ambiguity of the quantization.  The spin $l$ is introduced  in such a way,
that  for every spin-network state  $\Psi_\gamma$   the spin-network decomposition of the state $\epsilon_{ijk}(h^i_{\alpha_{ee'}})^{(l)}\hat{J}^j_{v,e}\hat{J}^k_{v,e'}\Psi_\gamma$   does not contain a component of the zero spin at a  segment of $e$ or $e'$ \cite{yang2015new}. In other words,
neither edge  $e$ nor edge $e'$ can even partially disappear as the effect of the action.   We achieve that goal by fixing 
\begin{equation}\label{eq:lcondition}
\begin{aligned} 
l\ &=\ \frac{1}{2}, \ \ \ \ \ \ \ {\rm provided}\ \ \ \ \ \ \ j_e,j_{e'}> \frac{1}{2}, \\
l\ &=\ \frac{3}{2}, \ \ \ \ \ \ \ {\rm provided}\ \ \ \ \ \ \ j_e=1/2,j_{e'}=1 \text{ or } j_{e'}=\frac{1}{2},j_e=1, \\
l\ &=\ 1, \ \ \ \ \ \ \ {\rm otherwise}.
\end{aligned}
\end{equation}

 \item ${}^{\rm kin}\hH^L_{v,ee'}$  does not change any given  graph $\gamma$, and even commutes with each of the operators
 $\hat{J}^i_{v,e}$, 
 \begin{equation} 
^{\rm kin}\hH^L_{ee'}:=\sqrt{\delta_{ii'}\left(\epsilon_{ijk}J_{v,e}^jJ_{v,e'}^k\right)\left(\epsilon_{i'j'k'}J_{v,e}^{j'}J_{v,e'}^{k'} \right)}\left(\frac{2\pi}{\alpha}-\pi+\arccos\left[\frac{\delta_{kl}J_{v,e}^kJ_{v,e'}^l}{\sqrt{\delta_{kk'}J_{v,e}^kJ_{v,e}^{k'}}\sqrt{\delta_{kk'}J_{v,e'}^kJ_{v,e'}^{k'}}}\right]\right),
\end{equation}
The factor $\alpha$ is arbitrary, representing another residual ambiguity of the quantization. 
\end{itemize}

Next, we pass the operator to  $\mathcal{H}_{\rm vtx}$, by the duality (\ref{eq:dualaction}),
\begin{equation}
 \hH'_v\eta(\Psi_\gamma)\ :=\ \eta(\Psi_{\gamma}){}^{\rm kin}\hH_v,
 \end{equation} 
for every subspace $\eta({\cal H}_\gamma)$. 
Finally, in the Hilbert space $\mathcal{H}_{\rm vtx}$ we turn it into a symmetric operator
\begin{equation}\label{eq:definehvcon}
  \h\ :=\ \frac{1}{2}\left(\hH'_v +\  (\hH'_v)^\dagger\right)\ = \   (1+\beta^2)\hH^L_v +
\frac{1}{2}\left(\hH'^E_v +\  (\hH'^E_v)^\dagger\right)\ =    (1+\beta^2)\hH^L_v +
 \hH^E_v
 \end{equation}
  If we considered $({}^{\rm kin}\hH_v)^\dagger$  and do symmetrization in the kinematical Hilbert space,
the resulting operator
would break the diffeomorphism invariance.

In order to implement (\ref{theoperator}), one has  to find a basis in ${\cal H}_{\rm vtx}$ that consists of eigenstates of  $  \h$, 
$$   \h |v,\lambda\rangle=\lambda |v,\lambda\rangle,$$
restrict the Hilbert space to the physical sector defined by the non-negative eigenvalues, and consider an operator  
\begin{equation}\label{theoperator'}\widehat{\sqrt{-2\sqrt{|\det E(x)|}C^{\rm gr}(x)}}\,|v,\lambda\rangle\ =\ \sum_{v\in \Sigma}\delta(v,x)\sqrt{\lambda}\,|v,\lambda\rangle .\end{equation}
For the time being,  we do not even know  a single non-trivial eigenstate of the operator $\h$, except for the subspaces $\eta({\cal H}_\gamma)$ given by graphs $\gamma$   that have no non-degenerate vertices. What we do in the next section is to consider  a simplest subspace of ${\cal H}_{\rm vtx}$, which contains states of non-degenerate vertices and  is preserved by the action of the 
operator $  \h$. We study the properties of the operator $  \h$ therein. We  prove that  it is self-adjoint.   Hence, that subspace
does admit an orthogonal  decomposition into the eigenstates of $  \h$. 

\section{Restriction to a subspace of ${\cal H}_{\rm vtx}$ preserved by $\h$}\label{se:two}
\subsection{The subspace}
In this section we construct a subspace $\mathcal{H}_{v}\subset\mathcal{H}_{\rm vtx}$ from a single loop with a kink at a point v (there is only one possibility, up to the diffeomorphisms), which is denoted by graph $\gamma_0$ as follows.
$$
\gamma_0:=\makeSymbol{\raisebox{0.\height}{\includegraphics[width=0.2\textwidth]{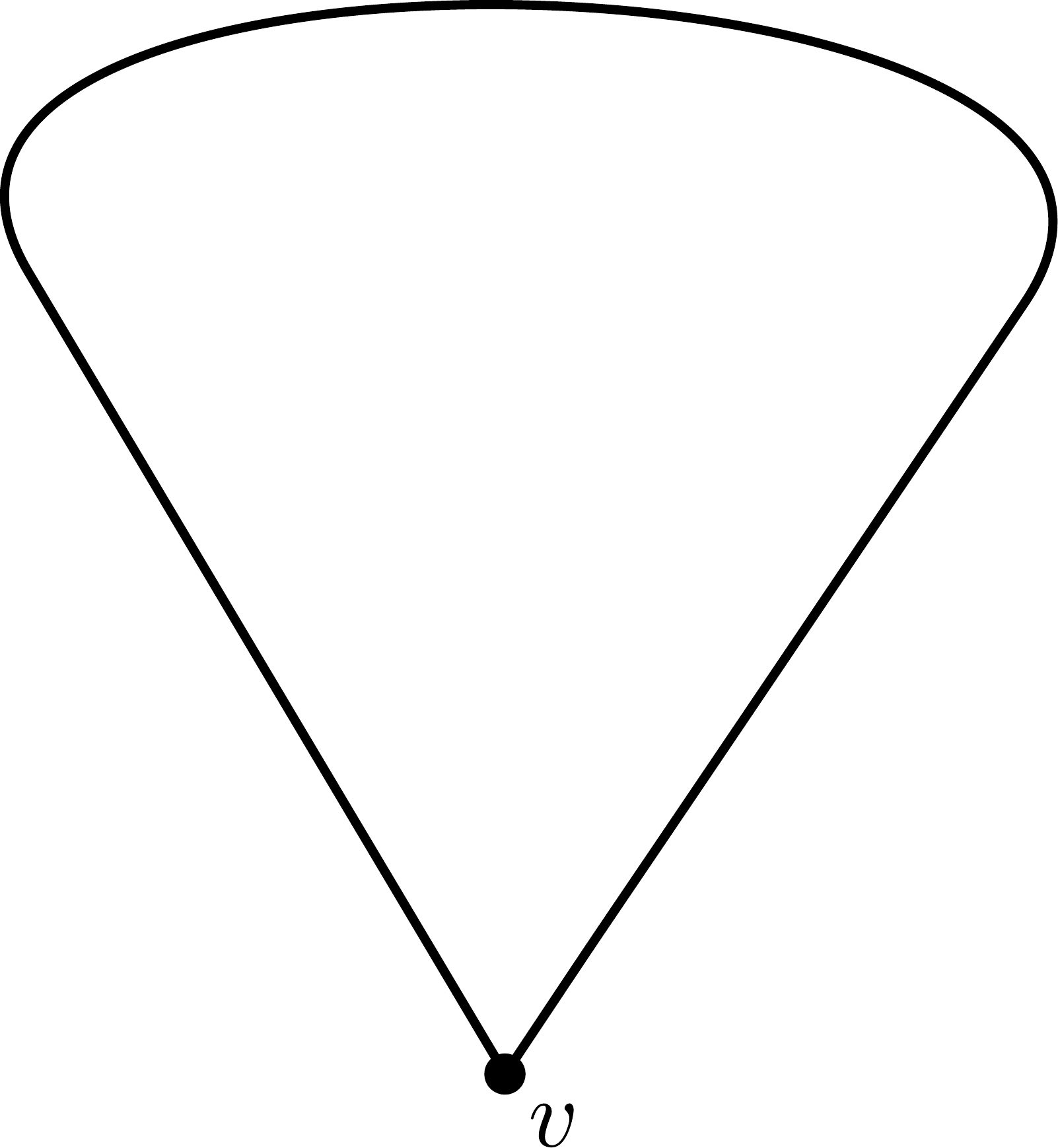}}}.
$$
The assumption that $\mathcal{H}_{v}$ contains the $\gamma_0$ and is preserved  by the operator $\h$  determines its construction. 
Fix a point $v\in \Sigma$. For every integer $n$ (including $n=0$) consider the following graph  $\gamma_n$ 
\begin{equation}\label{gamman}
\gamma_n:=\makeSymbol{\raisebox{0.\height}{\includegraphics[width=0.2\textwidth]{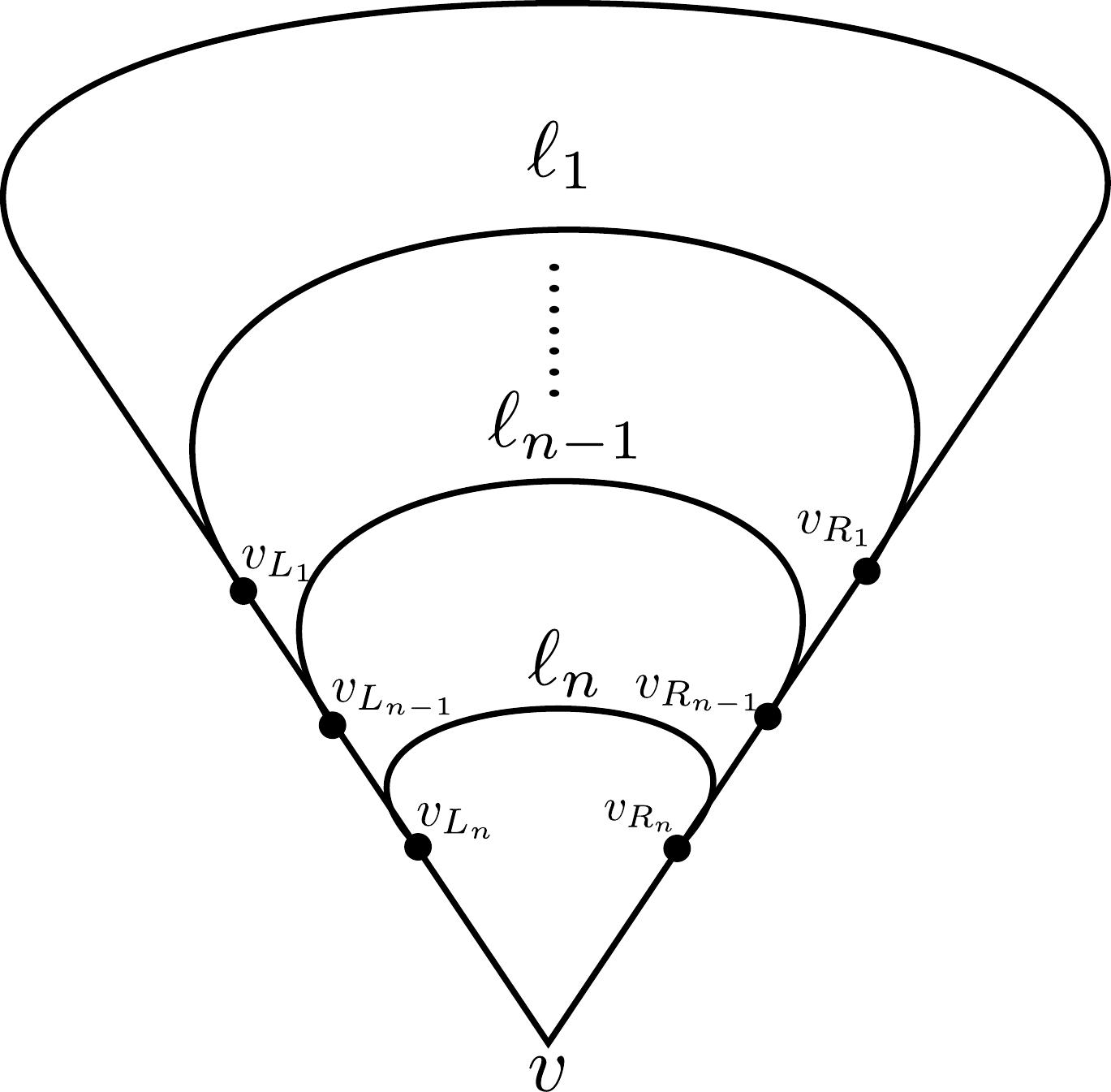}}}.
\end{equation}
The edges $\ell_1$, ..., $\ell_n$ are defined such that they could have been attached by the operator ${}^{\rm kin}\hH^E_v$
introduced above and acting  $n$ times in a row at the non-degenerate vertex $v$. The vertices $v_{L}{}_1, ... , v_{R}{}_n$ are 
ignored by the operator, because they are degenerate.  Each of the graphs defines a subspace, ${\cal H}_\gamma\in{\cal H}_{\rm kin}$
 of the kinematical Hilbert space \eqref{decompkin} and a subspace, $\eta({\cal H}_\gamma)\subset {\cal H}_{\rm vtx}$, of the space of diffeomorphism (preserving $v$) invariant  states. Consider the subspace ${\cal H}_\gamma^G$ of the Yang-Mills gauge \eqref{YM} 
invariant elements of ${\cal H}_\gamma $ and denote 
\begin{equation}
{\cal H}_{[\gamma_n]}\ :=\ \eta({\cal H}_{\gamma_n}^G). 
\end{equation}
The subspace $\mathcal{H}_v$ of ${\cal H}_{\rm vtx}$ we are constructing will be contained in the subspace 
\begin{equation}
\overline{\bigoplus_n {\cal H}_{[\gamma_n]}}\subset {\cal H}_{\rm vtx} .
\end{equation}

For every graph $\gamma_n$,  consider a spin-network state 
$|\gamma_n,\vec{j},\vec{l}~\rangle$, where $\vec{j}=(j_1,j_2,\cdots,j_{n+1})$ and $\vec{l}=(l_1,l_2,\cdots,l_n)$ are the spins 
assigned to edges of $\gamma_n$ as shown in the following equation (referring to Appendix \ref{se:appendix} for the graph notion)
\begin{equation}\label{eq:29fordiss}
 |\gamma_n,\vec{j},\vec{l}~\rangle:=\mathcal{N}_n\makeSymbol{\raisebox{0.\height}{\includegraphics[width=0.2\textwidth]{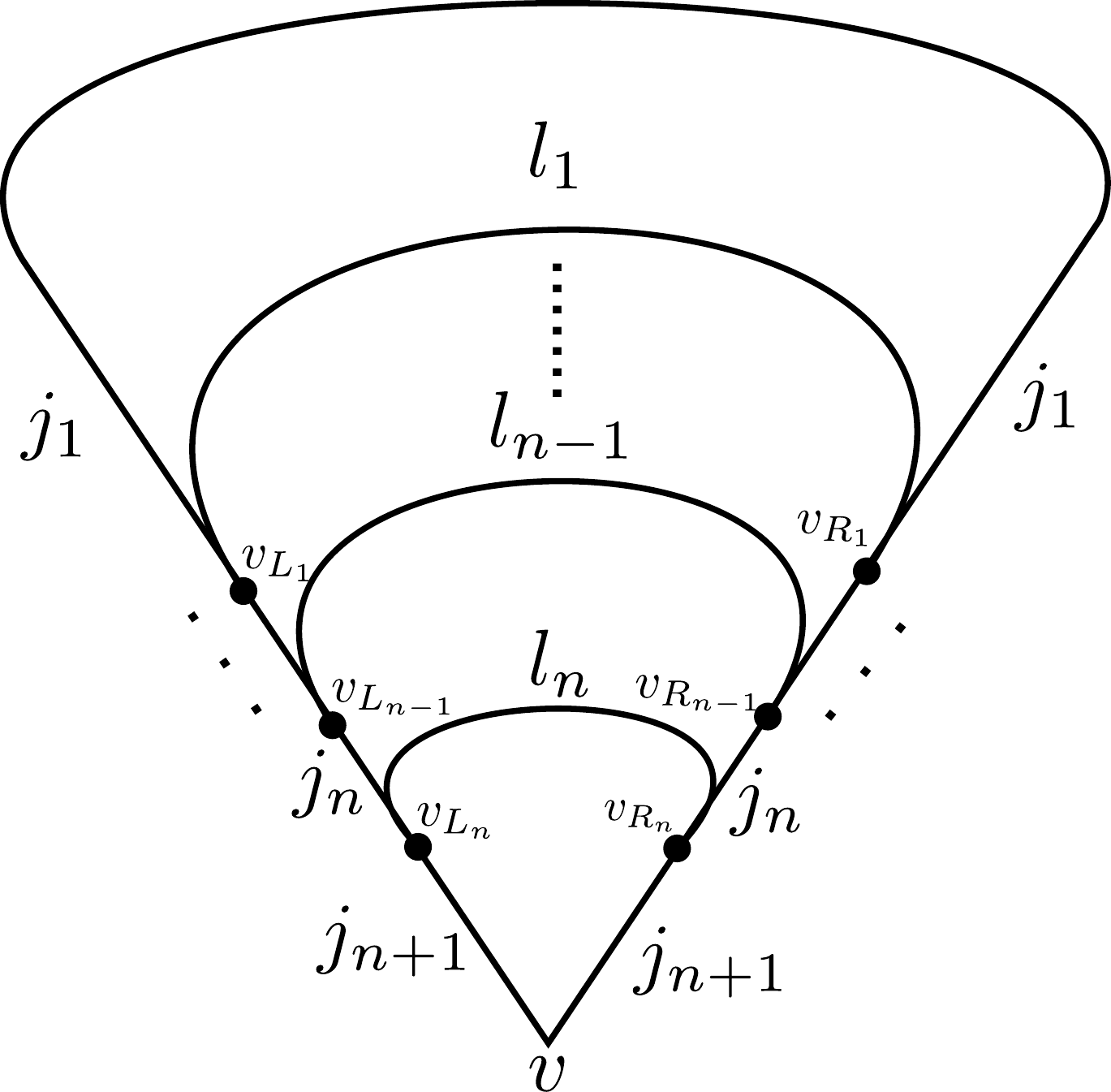}}}
\end{equation} 
where $\mathcal{N}$ is a real and positive normalization factor. If we act on any of those states $|\gamma_n,\vec{j},\vec{l}\rangle$ with operator $\h$, we obtain linear combination of  states $|\gamma_{n\pm 1},\vec{j}',\vec{l}'\rangle$ and $|\gamma_n,\vec{j},\vec{l}\rangle$ (see details below). Therefore, Hilbert space $\mathcal{H}_v$ is preserved by the action of operator $\h$. Hence we do not need to introduce other states.

Because all the vertices  of $\gamma_n$ are at most $3$-valent the  spin-network $ |\gamma_n,\vec{j},\vec{l}~\rangle$ is determined 
by the spins up to a phase factor. One can fix  the intertwiners to be the 3-$j$ symbols $
\left(\begin{matrix}
j_n&j_{n+1}&l_n\\
m_n&m_{n+1}&\mu_n
\end{matrix}\right)$ at 3-valent vertices and the 2-$j$ symbol $\epsilon^j_{mn}:=(-1)^{j+m}\delta_{m,-n}$ at the 2-valent vertex. The existence condition is that  $\vec{j}$ satisfies $|j_m-l_m|\leq j_{m+1}\leq j_m+l_m$, $m=1,2,\cdots,n$.  

The next step in the construction is application of the rigging map $\eta$ (\ref{eta}).  For every $n=0,1,...$ the corresponding graph 
$\gamma_n$ has a symmetry, $f_{\gamma_n}\in {\rm Diff}_v$, such that 
\begin{equation}\label{flip}
f_{\gamma_n}(v_L{}_k) = v_R{}_k, \ \ \ \ \ \ \ \ f_{\gamma_n}(v_R{}_k) = v_L{}_k . 
\end{equation}
Because of the symmetry of the graph $\gamma_n$, every $\Psi_{\gamma_n}\in{\cal H}_{\gamma_n}$ may has symmetric part $\psi_{\gamma_n}^+$ and antisymmetric part $\psi_{\gamma_n}^-$ with respect to the transformation \eqref{flip}, i.e. $\psi_{\gamma_n}^\pm=\pm \psi_{\gamma_n}^\pm$, while only the symmetric part contributes to  $\eta(\Psi_{\gamma_n})$.
We will show now, that each state $|\gamma_n,\vec{j},\vec{l}~\rangle$ defined above is  invariant with respect
to the symmetry (\ref{flip}). Consider  the function $\psi_{\gamma_n}$ (\ref{cyl}) corresponding to the state  
$|\gamma_n,\vec{j},\vec{l}~\rangle$,  
$$ \psi_{\gamma_n}({g}_L{}_1,...,{g}_L{}_{n+1},{g}_R{}_1,...,{g}_R{}_{n+1},h_1,...,h_n) \ =\ \langle \vec{g}_L\vec{g}_R\vec{h}|\gamma_n,\vec{j},\vec{l}~\rangle.$$ 
which is defined graphically as
\begin{equation}
\langle \vec{g}_L\vec{g}_R\vec{h}|\gamma_n,\vec{j},\vec{l}~\rangle=\mathcal{N}_n\makeSymbol{\raisebox{0.\height}{\includegraphics[width=0.4\textwidth]{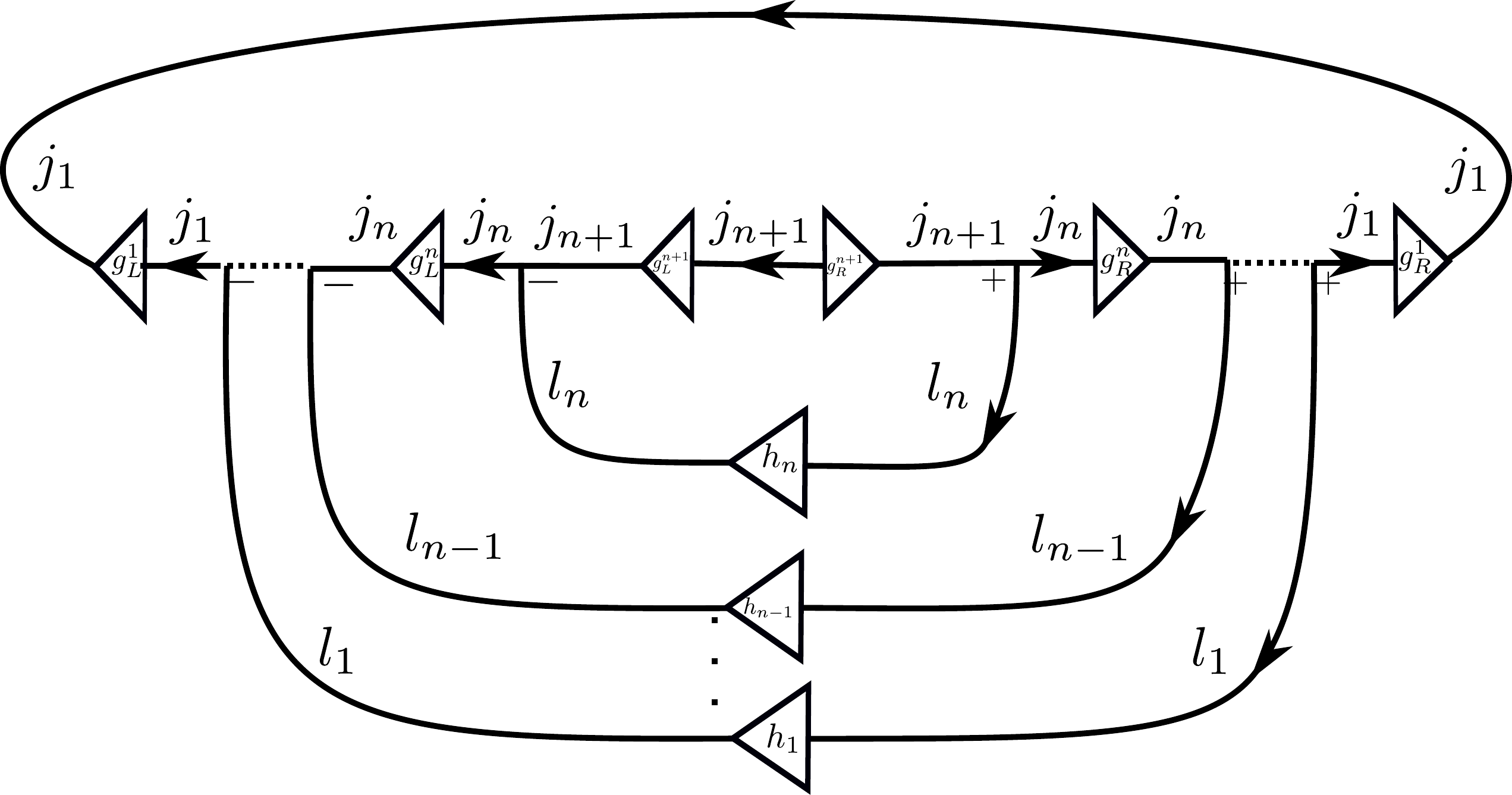}}}=:\mathcal{N}_n~\makeSymbol{\raisebox{0.\height}{\includegraphics[width=0.3\textwidth]{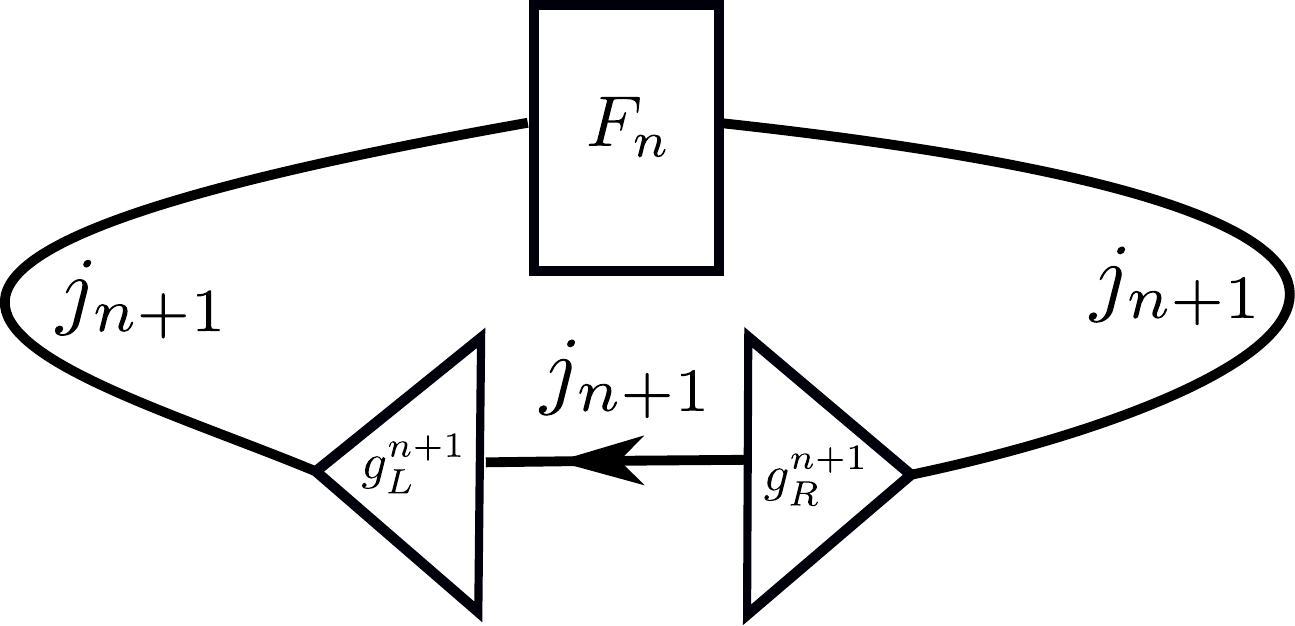}}}.
\end{equation}
Then, the cylindrical function $\psi'_n$ corresponding to the flipped state $U_{f_{\gamma_n}}|\vec{j},\vec{l}~\rangle$ 
is
$$\psi'_{\gamma_n}({g}_L{}_1,...,{g}_L{}_{n+1},{g}_R{}_1,...,{g}_R{}_{n+1},h_1,...,h_n) \ =\ \psi_{\gamma_n}({g}_R{}_1,...,{g}_R{}_{n+1},{g}_L{}_1,...,{g}_L{}_{n+1},(h_1)^{-1},...,(h_n)^{-1}) . $$
We show below that indeed, 
$$\psi_{\gamma_n}({g}_R{}_1,...,{g}_R{}_{n+1},{g}_L{}_1,...,{g}_L{}_{n+1},(h_1)^{-1},...,(h_n)^{-1})\ =\  
\psi_{\gamma_n}({g}_L{}_1,...,{g}_L{}_{n+1},{g}_R{}_1,...,{g}_L{}_{n+1},h_1,...,h_n).$$
Using the following 
properties,
$$\left(
\begin{matrix}
j_1&j_2&j_3\\
m_1&m_2&m_3
\end{matrix}
\right)=\left(
\begin{matrix}
j_3&j_1&j_2\\
m_3&m_1&m_2
\end{matrix}
\right)=(-1)^{j_1+j_2+j_3}\left(
\begin{matrix}
j_3&j_2&j_1\\
m_3&m_2&m_1
\end{matrix}
\right), \text{ and, }\epsilon^j_{mn}=(-1)^{2j}\epsilon^j_{nm},$$
we can get
\begin{equation}
\langle \vec{g}_L\vec{g}_R\vec{h}|\hat{U}_{f_{\gamma_n}}|\gamma_n,\vec{j},\vec{l}~\rangle=(-1)^{2j_1+2l_1+2l_2+\cdot+2l_n+2j_{n+1}}\langle \vec{g}_L\vec{g}_R\vec{h}|\gamma_n,\vec{j},\vec{l}~\rangle.
\end{equation}
Because $j_m+l_m+j_{m+1}$ is integer, we have
\begin{equation}
(-1)^{2j_1+2l_1+2l_2+\cdot+2l_n+2j_{n+1}}=(-1)^{2j_2+2l_2+\cdots+2l_n+2j_{n+1}}=\cdots=(-1)^{4j_{n+1}}
=1.
\end{equation}  
Therefore
\begin{equation}
\langle \vec{g}_L\vec{g}_R\vec{h}|\hat{U}_{f_{\gamma_n}}|\gamma_n,\vec{j},\vec{l}~\rangle=\langle \vec{g}_L\vec{g}_R\vec{h}|\gamma_n,\vec{j},\vec{l}~\rangle .
\end{equation}
This ensures that $\langle \vec{g}_L\vec{g}_R\vec{h}|\gamma_n,\vec{j},\vec{l}~\rangle$ will not be  annihilated by the rigging map $\eta$.
We  denote
\begin{equation}
 \dualvec{[\gamma_n],\vec{j},\vec{l}~}:=\eta\big(|\gamma_n,\vec{j},\vec{l}~\rangle\big).
 \end{equation} 
We  further restrict the set of states that will define the subspace ${\cal H}_v$, by adjusting the spins $l_1,...l_n$ to the spins $j_1,...,j_n$ in the way corresponding to the action of the operator ${}^{\rm kin}\hH^E$ of  \eqref{eq:eucleadianH}. Hence, $l_n$ becomes a function of $j_{n}$ as that in \eqref{eq:lcondition}. It is simplified in our case as
\begin{equation}\label{eq:lofj}
l_n\ =\ l(j_n)\ =\ \left\{
\begin{matrix}
1&,\ j_n=1/2,\\
1/2&,\ j_n\neq 1/2.
\end{matrix}
\right.
\end{equation}
Then $\dualvec{[\gamma_n],\vec{j},\vec{l}~}$ can be abbreviated to $\dualvec{[\gamma_n],\vec{j}~}$. Finally, 
the Hilbert space $\mathcal{H}_v$ is defined as 
\begin{equation}\label{eq:defhvcon}
\mathcal{H}_v\ =\ \overline{{\rm Span}\{\dualvec{[\gamma_n],\vec{j}}~{\rm with }~|j_m-l_m(j_m)|\leq j_{m+1}\leq j_m+l_m(j_m),\forall m\leq n\}}.
\end{equation}
The natural domain $\mathcal{F}\ \subset\ \mathcal{H}_v$  for our operators will be the space of the finite linear combinations of the states $\left([\gamma_n],\vec{j}~\right|$, 
\begin{align}\label{eq:F}
\mathcal{F}\ :=\ {\rm Span}\{\dualvec{[\gamma_n],\vec{j}}~{\rm with }~|j_m-l_m(j_m)|\leq j_{m+1}\leq j_m+l_m(j_m),\forall m\leq n\}.
\end{align}

\subsection{The action of the operator $\h$ on $\mathcal{H}_v$}\label{se:definitionhE}
We calculate now the action of the operator $\h$ defined in the section \ref{se:phyH} on the space ${\cal H}_v$. 
{Following the framework,  we start with the   operators ${}^{\rm kin}\hH_v$,  ${}^{\rm kin}\hH^L_{ee'}$ and ${}^{\rm kin}\hH^E_{ee'}$.   For every   graph $\gamma_n$ of \eqref{gamman} the vertex $v$    is the only  non-degenerate vertex  such that
${}^{\rm kin}\hH_v\not= 0$,  and there is only one pair of edges meeting at $v$ and featuring in (\ref{eq:physicalH}). They are
$$(e, e') = (e_L,e_R)$$
connecting the vertex $v$ with the vertex $v_L{}_n$ and with  the vertex $v_R{}_n$, respectively. 
  Therefore
\begin{equation}\label{eq:physicalHv}
{}^{\rm kin}\h =\   {}^{\rm kin}\hH_{v,e_Le_R}\ =\   (1+\beta^2){}^{\rm kin}\hH^L_{v,e_Le_R}+{}^{\rm kin}\hH^E_{v,e_Le_R}.
\end{equation}
 For the Lorentz part, which is shown to be diagonalized under the basis, we have
\begin{equation}
{}^{\rm kin}\hH^L_v|\gamma_n,\vec{j},\vec{l}~\rangle \  =\ \frac{\pi}{\alpha}\sqrt{j_{n+1}(j_{n+1}+1)}|\gamma_n,\vec{j},\vec{l}~\rangle.
\end{equation}
That formula passes to the dual states, elements of the Hilbert space ${\cal H}_v$, simply as
\begin{equation}\label{HLfinal}
\hH^L_v\dualvec{[\gamma_n],\vec{j}}=\dualvec{[\gamma_n],\vec{j}}{}^{\rm kin}\hH^L_v=\frac{\pi}{\alpha}\sqrt{j_{n+1}(j_{n+1}+1)}\dualvec{[\gamma_n],\vec{j}}.
\end{equation}
The Euclidean part   (\ref{eq:eucleadianH}) is more complicated.   The  straightforward calculation in Appendix \ref{app:calculation} shows that }
\begin{equation}\label{eq:result1}
\begin{aligned}
&{}^{\rm kin}\hH^E_v|\gamma_n,\vec{j},\vec{l}\rangle\\
=& \frac{-3}{l_{n+1}(l_{n+1}+1)(2l_{n+1}+1)} \kappa_1\sum_{j_{n+2}} \frac{\sqrt{2j_{n+2}+1}}
{\sqrt{(2j_{n+1}+1)(2l_{n+1}+1)}}\left(\vec{J}_{n+1}\cdot\vec{L}_{n+1}\right)\left|\gamma_{n+1},(\vec{j},j_{n+2}),(\vec{l},l_{n+1})\right\rangle,
\end{aligned}
\end{equation}
  where we have denoted (see \eqref{J})
\begin{align} 
\vec{J}_{n+1}\ &=\ (J^1_{v_L{}_n,e_L},J^2_{v_L{}_n,e_L}, J^3_{v_L{}_n,e_L}),\\
\vec{L}_{n+1}\ &=\ (J^1_{v_L{}_n,\ell_{n+1}},J^2_{v_L{}_n,\ell_{n+1}}, J^3_{v_L{}_n,\ell_{n+1}}),
\end{align}
and given $\vec{j}=(j_1,...,j_{n+1})$ the symbol $(\vec{j}, j_{n+2})$ standing for $(j_1,...,j_{n+1},j_{n+2})$.
 Because of  \eqref{eq:lofj}, we can explicitly show out the factor
$$
\frac{-3}{l_{n+1}(l_{n+1}+1)(2l_{n+1}+1)}=\left\{
\begin{matrix}
-2,~j_{n+1}\geq 1\\
-\frac{1}{2},~j_{n+1}=\frac{1}{2}
\end{matrix}
\right.=:\omega(j_{n+1}).
$$
By defining the following function
\begin{equation}
\eta(x):=\left\{
\begin{matrix}
1,& x<1,\\
x,& x\geq 1.
\end{matrix}
\right.
\end{equation}
it follows from Eq. \eqref{eq:lofj} immediately that
\begin{equation}
j_{n+1}=\eta(j_n)\pm \frac{1}{2}.
\end{equation}
In the following, we frequently use $\pm$ instead of $\eta(j_n)\pm 1/2$ in the following. For example, we may write $(j_1,j_2,\cdots,j_n,-)$ rather than $(j_1,j_2,\cdots,j_n,\eta(j_n)-1/2)$, and $(j_1,-,+,\cdots,)$ can also be used instead of $(j_1,\eta(j_1)-1/2,\eta(\eta(j_1)-1/2)+1/2,\cdots)$. Then \eqref{eq:result1} can be rewritten as
\begin{equation}\label{eq:allj}
\begin{aligned}
{}^{\rm kin}\hH^E|\gamma_n,\vec{j}\rangle= \kappa_1\omega(j_{n+1})\left(\frac{\eta(j_{n+1})\sqrt{\eta(j_{n+1})+1}}{2\sqrt{2\eta(j_{n+1})+1}}
\left|\gamma_{n+1},(\vec{j},+)\right\rangle- \frac{[\eta(j_{n+1})+1]\sqrt{\eta(j_{n+1})}}{2\sqrt{2\eta(j_{n+1})+1}}
\left|\gamma_{n+1},(\vec{j},-)\right\rangle\right).
\end{aligned}
\end{equation}
By the definition, we have
\begin{equation}
\hH'^E_v\cdot\dualvec{[\gamma_n],\vec{j}}=\dualvec{[\gamma_n],\vec{j}}~{}^{\rm kin}\hH^E_v= \kappa_1 \frac{\sqrt{\eta(j_n)(\eta(j_n)+1)}}{2\sqrt{2\eta(j_n)+1}}\Theta(j_{n+1},j_n)~ \dualvec{[\gamma_{n-1}],(j_1,j_2,\cdots,j_n)},
\end{equation}
where 
\begin{equation}\label{eq:Theta}
\Theta(j_{n+1},j_n)=\left\{\begin{matrix}
\omega(j_{n})\sqrt{\eta(j_{n})}&, \ \ {\rm if}\ \ j_{n+1}=\eta(j_n)+1/2,\\
 ~\\
-\omega(j_{n})\sqrt{\eta(j_n)+1}&, \ \ {\rm if}\ \ j_{n+1}=\eta(j_n)-1/2.
\end{matrix} \right.
\end{equation}
  Hence, in our subspace ${\cal H}_v$, the operator $\hH'^E_v$ just annihilates 
the edges $\ell_{n}$ of the graphs $\gamma_n$.  On  the other hand, the adjoint operator $(\hH'^E_v)^\dagger$ acts by
creating new edges $\ell$  by a formula very similar to that of
${}^{\rm kin}\hH^E_v$,
namely
\begin{equation}\label{eq:hdagger}
\begin{aligned}
& (\hH'^E_v)^\dagger\cdot\dualvec{ [\gamma_n],\vec{j}}\\
=& \kappa_1\omega(j_{n+1})\frac{\eta(j_{n+1})\sqrt{\eta(j_{n+1})+1}}{2\sqrt{2\eta(j_{n+1})+1}}
\dualvec{[\gamma_{n+1}],(\vec{j},+)}- \kappa_1\omega(j_{n+1})\frac{[\eta(j_{n+1})+1]\sqrt{\eta(j_{n+1})}}{2\sqrt{2\eta(j_{n+1})+1}}
\dualvec{[\gamma_{n+1}],(\vec{j},-)}.
\end{aligned}
\end{equation}
  Finally,  we obtain a formula for the symmetric part, and the action of the operator  $\hH^E_v$ in  ${\cal H}_v$,
\begin{equation}\label{eq:definitionofhhd}
\begin{aligned}\
&\dualvec{[\gamma_n],\vec{j}} \hH^E_v=\frac{1}{2}(\hH'^E_v+(\hH'^E_v)^\dagger)\cdot\dualvec{[\gamma_n],\vec{j}}\\
=&  \frac{\kappa_1}{2}\sum_{j_{n+2}=\eta(j_{n+1})\pm 1/2} \zeta(j_{n+1})
\Theta(j_{n+2},j_{n+1})\dualvec{[\gamma_{n+1}],(\vec{j},j_{n+2})}+ \zeta(j_{n})\Theta(j_{n+1},j_n) \dualvec{[\gamma_{n-1}],(j_1,j_2,\cdots,j_n)},
\end{aligned}
\end{equation}
where
\begin{equation}\label{eq:zeta}
\zeta(j):=\frac{\sqrt{\eta(j)(\eta(j)+1)}}{2\sqrt{2\eta(j)+1}}.
\end{equation}

  In summary, we have given the action of the operator $H_v$ defined in the domain ${\cal F}$ of the subspace ${\cal H}_v$ of the vertex Hilbert space ${\cal H}_{\rm vtx}$. It can be repressed as 
\begin{equation}\label{eq:finalhv}
\h\ =\ (1+\beta^2)\hH^L_v + \hH^E_v,
\end{equation}
where the terms of the right hand side are defined in (\ref{HLfinal}) and (\ref{eq:definitionofhhd}).
There are present several arbitrary constant factors. 
The first term involving $\hH^L_v$ is proportional to a positive constant factor  $\frac{1+\beta^2}{\alpha}$, while the second one to a positive constant factor $\kappa_1$. The factors represent ambiguity of the quantization \cite{alesci2015hamiltonian}. For the analysis of the problem of self-adjointness of the operator we can fix one of those factors arbitrarily. Hence we set
$$\kappa_1\ =\ 2. $$

{ 
\section{Self-adjointness of the operators}\label{se:three}
In this section we will prove the following result:
\medskip

\begin{theorem}\label{theorem}
On the Hilbert space $\mathcal{H}_v$, the operator $\h$ defined by \eqref{eq:finalhv} in Sec. \ref{se:definitionhE} in the domain ${\cal F}$ is essentially self-adjoint. 
\end{theorem}
\medskip

First, we sketch the proof. 
Consider the following operator $\z$ defined  in ${\cal F}$, 
\begin{equation}
\dualvec{[\gamma_n],\vec{j}}\z: =\zeta(j_{n+1})\dualvec{[\gamma_n],\vec{j}},
\end{equation}
and introduce the operator
$$ \n\ :=\ \z^2.$$
By definition, $\bar{\n}$, the closure of $\n$, is self-adjoint. $\f$ is a core of $D(\bar{\n})$. 

The key part of the proof is the following Lemma:
\smallskip

\begin{lemma}\label{lemma} There exist $c,d\in \mathbb{R}^+$ such that for every $\dualvec{\psi}\in{\cal F}$ the following
two inequalities are true:
\begin{equation}
\begin{aligned}\label{theineq}
\left|\left|\dualvec{\psi}\h\right|\right|^2&\leq c \left|\left|\dualvec{\psi}\n\right|\right|^2, \\
\left|\dualvec{\psi}[\h,\n]\Big|\psi\Big)\right|&\leq d\left|\left|\dualvec{\psi}\n^{1/2}\right|\right|^2.
\end{aligned}
\end{equation}
\end{lemma} 
\smallskip
It turns out (see Appendix C) that  Theorem \ref{theorem} follows directly from Lemma \ref{lemma}.

Note that in the calculations proving Lemma  \ref{lemma} it is convenient to express the Euclidean part $\hH^E_v$
of the operator $\hH_v$ by the following "creation" and "annihilation" operators $\ada$ and $\haa$, respectively:
 
\begin{equation}
\dualvec{[\gamma_n],\vec{j}}\haa\ :=\ \Theta(j_{n+1},j_n)\dualvec{[\gamma_{n-1}],(j_1,\cdots,j_n)}
\end{equation}
\begin{equation}
\dualvec{[\gamma_n],\vec{j}}\ada\ :=\ \sum_{j_{n+2}=\eta(j_{n+1}\pm 1/2)}\Theta(j_{n+2},j_{n+1})\dualvec{[\gamma_{n+1}],(\vec{j},j_{n+2})}.
\end{equation}
Of course $\ada$ is adjoint to $a$ and restricted to $\f$. 
In terms of the operators $\z$, $\haa$ and $\ada$, we have
\begin{equation}
\hH^E_v =\haa\z+\z\ada.
\end{equation}

Now we come back to the proof of the lemma \ref{lemma}.
Given $\dualvec{\psi}=\sum_{n,\vec{j}}\beta_{n,\vec{j}}\dualvec{[\gamma_n],\vec{j}~}$, we have
\begin{equation*}
\dualvec{\psi}\h=\sum_{m,\vec{i}}\dualvec{[\gamma_m],\vec{i}}\sum_{n,\vec{j}}\beta_{n,j}\dualvec{[\gamma_n],\vec{j}}\h\Big|[\gamma_m],\vec{i}\Big)
\end{equation*}
which gives us
\begin{equation}\label{eq:gamehv}
\begin{aligned}
||\dualvec{\psi}\h||^2&=\sum_{m,\vec{i}}\left|\sum_{n,\vec{j}}\dualvec{[\gamma_n],\vec{j}}(\h)\Big|[\gamma_m],\vec{i}\Big)\beta_{n,\vec{j}}\right|^2\\
&\leq 4\sum_{m,\vec{i}}\sum_{n,\vec{j}}\left|\dualvec{[\gamma_n],\vec{j}}\h\Big|[\gamma_m],\vec{i}\Big)\right|^2|\beta_{n,\vec{j}}|^2\\
&= 4\sum_{n,\vec{j}}\left(\sum_{m,\vec{i}}\left|\dualvec{[\gamma_n],\vec{j}}\h\Big|[\gamma_m],\vec{i}\Big)\right|^2\right)|\beta_{n,\vec{j}}|^2,
\end{aligned}
\end{equation}
where the factor of $4$ is due to the fact that there are only 4 non-vanishing entries in each row of the matrix of $\h$. 
By \eqref{HLfinal} and \eqref{eq:definitionofhhd}, we get

\begin{equation*}
\begin{aligned}
&\sum_{m,\vec{i}}\left|\dualvec{[\gamma_n],\vec{j}}\h\Big|[\gamma_m],\vec{i}\Big)\right|^2\\
=&(1+\beta^2)\frac{\pi}{\alpha}j_{n+1}(j_{n+1}+1)+\sum_{j_{n+2}=\eta(j_{n+1})\pm 1/2}[ \zeta(j_{n+1})
\Theta(j_{n+2},j_{n+1})]^2+ [\zeta(j_{n})\Theta(j_{n+1},j_n) ]^2
\end{aligned}
\end{equation*}
When $j_{n+1}\to \infty$, the right hand side, as well as $\zeta(j_{n+1})^4$, increases as $\sim j_{n+1}^2$. Hence there must exist a number $c\in \mathbb{R}^+$ such that 
$$
\sum_{m,\vec{i}}\left|\dualvec{[\gamma_n],\vec{j}}\h\Big|[\gamma_m],\vec{i}\Big)\right|^2\leq c\zeta(j_{n+1})^4,
$$
for all $j_{n+1}\geq \frac{1}{2}$. Therefore we have
\begin{equation}\label{eq:condition1}
||\h\psi||^2\leq 4c\sum_{n,\vec{j}}\zeta(j_{n+1})^4|\beta_{n,\vec{j}}|^2=4c\sum_{n,\vec{j}}|\zeta(j_{n+1})^2\beta_{n,\vec{j}}|^2=4c||N\psi||^2.
\end{equation}

For the second equation in \eqref{theineq}, we define 
\begin{equation*}
\hat{C}:=i\frac{1}{\z}[\h,\n]\frac{1}{\z}=i(\frac{1}{\z}\haa\z^2-\z \haa+\ada \z-\z^2 \ada\frac{1}{\z}).
\end{equation*}
Playing the same game as what we did for $\h$ in \eqref{eq:gamehv}, we obtain
\begin{equation*}
\begin{aligned}
\left|\left|\dualvec{\psi}\hat{C}\right|\right|^2\leq 3\sum_{n,\vec{j}}\left(\sum_{m,\vec{i}}\left|\dualvec{ [\gamma_n],\vec{j}}\hat{C}\Big|[\gamma_m],\vec{i}\Big)\right|^2\right)|\beta_{n,\vec{j}}|^2,
\end{aligned}
\end{equation*}
Because of
\begin{equation*}
\begin{aligned}
\dualvec{[\gamma_n],\vec{j}}\hat{C}=&i\sum_{j_{n+2}=\eta(j_{n+1})\pm 1/2}\Theta(j_{n+2},j_{n+1})\left(\frac{\zeta(j_{n+1})^2}{\zeta(j_{n+2})}-\zeta(j_{n+2})\right) \dualvec{[\gamma_{n+1}],(\vec{j},j_{n+2})}\\
+&i\Theta(j_{n+1},j_{n})\left(\zeta(j_{n+1})-\frac{\zeta(j_{n+1})}{\zeta(j_n)}\right) \dualvec{[\gamma_{n-1}],(j_1,\cdots,j_n)},
\end{aligned}
\end{equation*}
we have
\begin{equation*}
\begin{aligned}
&\sum_{m,\vec{i}}\left|\dualvec{[\gamma_n],\vec{j}}\hat{C}\Big|[\gamma_m],\vec{i}\Big)\right|^2\\
=&\sum_{j_{n+2}=\pm}\left(\frac{\zeta(j_{n+1})^2}{\zeta(j_{n+2})}-\zeta(j_{n+2})\right)^2\Theta(j_{n+2})^2+\left(\zeta(j_{n+1})-\frac{\zeta(j_{n})}{\zeta(j_{n+1})}\right)^2\Theta(j_{n+1})^2.
\end{aligned}
\end{equation*}
It is easy to check that the function
$$
\left(\frac{\zeta(j)^2}{\zeta(\eta(j)\pm 1/2)}-\zeta(\eta(j)\pm 1/2)\right)^2\Theta(\eta(j)\pm 1/2,j)
$$
is bounded for $j\geq 0$. Hence there exists a $d>0$ such that
\begin{equation*}
\left|\left|\dualvec{\psi}\hat{C}\right|\right|^2\leq d||\psi||^2,
\end{equation*}
which means that $\hat{C}$ is bounded. Because $\dualvec{\psi}\z\in\f$ is well defined for all $\dualvec{\psi}\in \f$, we obtain
\begin{equation}\label{eq:condition2}
\left|\dualvec{\psi}[\h,\n]\Big|\psi\Big)\right|=\left|\dualvec{\psi}\z \hat{C}\z\Big|\psi\Big)\right|\leq ||\hat{C}||~ \left|\left|\dualvec{\psi}\z\right|\right|^2=||\hat{C}||~ \left|\left|\dualvec{\psi}\n^{1/2}\right|\right|^2,~\forall \dualvec{\psi}\in \f.
\end{equation}
This completes the proof of the lemma \ref{lemma}.
  In conclusion, according to  Theorem \ref{theom:commutator-SA} in Appendix \ref{sec:C} and the Lemma \ref{lemma}, the operator $H_v$ defined in the domain $\f$ is essentially self-adjoint on ${\cal H}_v$.  Because the above proof is valid also for the case when $(1+\beta^2)=0$, the Euclidean part $\hH^E_v$ of $\h$ is also essentially self-adjoint by itself. 
 \section{Eigenvalue problem}\label{se:four}
Let $\dualvec{\psi}=\sum_{n,\vec{j}}\dualvec{[\gamma_n]\vec{j}}\psi_{n,\vec{j}}$ be an eigenstate of $\h$ with the eigenvalue $\lambda$. By definition we have
\begin{equation}\label{eq:recurencerelation}
\begin{aligned}
&\sum_{j_{n+2}=\eta(j_{n+1})\pm 1/2} \zeta(j_{n+1})
\Theta(j_{n+2},j_{n+1})\psi_{n+1,(\vec{j},j_{n+2})}+\zeta(j_{n})\Theta(j_{n+1},j_n) \psi_{j_{n-1},(j_1,\cdots,j_n)} \\
+&\frac{\pi(1+\beta^2)}{\alpha}\sqrt{j_{n+1}(j_{n+1}+1)}\psi_{n,\vec{j}}=\lambda\psi_{n,\vec{j}}. 
\end{aligned}
\end{equation}
To understand the recurrence equations for all coefficients $\psi_{n,\vec{j}}$, we introduce a  triangle array of  the coefficients of $\dualvec{\psi}$  as follows.  In the array, the rows are conventionally enumerated starting with $n=0$ at the top. There are $2^n$ entries in the nth row. The entries in the nth row are the coefficients with index $n$ (i.e. $\dualvec{\psi}[\gamma_n],\vec{j}\Big)$ for various $\vec{j}$~). The coefficients $\dualvec{\psi}[\gamma_{n+1}],(\vec{j},\eta(j_{n+1})\pm 1/2)\Big)$ are listed below to the left and right of $\dualvec{\psi}[\gamma_n],\vec{j}\Big)$, i.e.,
$$
\begin{tabular}{*{3}{c}}
 &$\psi_{n,\vec{j}}$\\
 $\psi_{n+1,(\vec{j},-)}$&&$\psi_{n+1,(\vec{j},+)}$
\end{tabular}
$$
We call the array as coefficient triangle.
In a given recurrence equation \eqref{eq:recurencerelation}, involved block looks like\footnote{By definition, $\psi_{n,\vec{j}}$ in the block should be located below to the left or right depending on the sign of $j_{n+1}$. } 
$$
\begin{tabular}{*{3}{c}}
 &$ \psi_{n-1,(j_1,\cdots,j_n)}$\\
 &$\psi_{n,\vec{j}}$\\
 $ \psi_{n+1,(\vec{j},-)}$&&$\psi_{n+1,(\vec{j},+)}$
\end{tabular}
$$
The two coefficients $\psi_{n-1,(j_1,j_2,\cdots,j_n)}$ and $\psi_{n+1,\vec{j}}$ on the top are either fixed, or derived by previous recurrence equations. Thus, for instance, in order to determine $\psi_{n+1,(\vec{j},+)}$ by \eqref{eq:recurencerelation}, the other coefficient $\psi_{n+1,(\vec{j},-)}$ in the same row must be fixed by hand.  It follows immediately that $1+\sum_{n=0}^\infty 2^n$ initial data should be fixed to solve all the recurrence equations in \eqref{eq:recurencerelation}. One choice of the initial data, for instance, is to fix $\psi_{0,j_1}$ and $\psi_{n,(j_1,\cdots,j_n,-)}$. The degrees of degeneracy of eigenstates for a given eigenvalue is therefore $\sum_{n=0}^\infty 2^n$. We leave the resolution of this complicated eigenvalue problem for further study.
\section{Summary and discussion} \label{se:five}
The  general issue addressed in our paper is  to understand the analytic properties of  Hamiltonian operators of LQG that follow from attaching and removing loops. Particularly, the interesting  case for us is 
when the action of the Hamiltonian operator could not be reduced to a single, finite graph.  We considered the effective Hamiltonian operator of  LQG coupled to the massless Klein-Gordon field. The model has  infinitely many local degrees of freedom.  To simplify the task, we restricted our analysis to the effects on the properties of the operator produced by the creation and annihilation of the loops at a single pair of edges intersecting at a vertex $v$. 
Therefore, we constructed a smallest subspace $\mathcal{H}_v$ \eqref{eq:defhvcon} of the vertex Hilbert space $\mathcal{H}_{\rm vtx}$,  which has the following desired properties: $(i)$ being preserved by the quantum Hamiltonian operator, and  $(ii)$  containing  a spin-network state defined by the graph $\gamma_0$   depicted at the beginning of Sec. \ref{se:two} colored by a non-trivial representation $j_1$.  The subspace is still infinite dimensional, and for arbitrary integer $n$ it contains a spin-network state \eqref{eq:29fordiss}.  It properly captures the properties of the Hamiltonian operator we wanted to know. We have restricted our study  to this subspace $\mathcal{H}_v$.  Therein,
we considered the operator $\h$ defined by \eqref{eq:definehvcon}, which was employed in the construction of the physical Hamiltonian operator \eqref{theoperator}. The action  of $\h$ in $\mathcal{H}_v$ is analysed in detail, and the explicit formulae for its matrix elements in a suitable normalized basis $\dualvec{[\gamma_n],\vec{j}}$ are derived. It turns out that the operator $\h$ possess the following properties, which are relevant for further analysis,
\begin{itemize}
\item each state $\dualvec{[\gamma_n],\vec{j}}$ is mapped by $H_v$ into a linear combination of at most $4$ elements of that basis; 
\item the coefficients depend on $j_{n+1}$ only, and are of the order of $j_{n+1}$ in the limit $j_{n+1}\rightarrow \infty$.  
\end{itemize}    
These properties are crucially used in the proof of the lemma \ref{lemma}, which ensure the self-adjointness of $\h$ on $\mathcal{H}_v$. Since $\h$ is self-adjoint, the physical Hamiltonian operator $\hat{H}_{\rm phy}:=\sqrt{\h}$ can be well defined in $\mathcal{H}_v$ by restricting to the non-negative part of the spectrum of $\h$.  Moreover, our analysis  gives insight into the eigenvector problem of $\hat{H}_v$. However
we have not found a normalizable solution.

It is desirable to further generalize the above result of $\h$ to the vertex Hilbert space $\mathcal{H}_{\rm vtx}$. If the matrix elements of $\h$ increased linearly with the spins like the second property above, the Theorem \ref{theorem} could still be employed for the generalization. However, a tentative calculation shows that it is not the case for vertices of arbitrary valency. For instance, some quadratic terms of spins will appear in the case of 3-valent vertices. Thus, the generalization of our result to $\mathcal{H}_{\rm vtx}$ is not straight forward. This issue has to be left for further study. 

\section{Acknowledgement}
We are benefited greatly from the numerous discussions with Wojciech Kaminski and Ilkka M\"akinen. 
Cong Zhang gratefully acknowledges financial support from China Scholarship Council(CSC), NO. 201606040084.
This work was
supported by  the  Polish Narodowe Centrum Nauki, Grant No. 2011/02/A/ST2/00300 and the Natural Science Foundation of China (NSFC), grand No. 11475023 and No. 11875006.
\appendix
\section{Graphical Calculation method}\label{se:appendix}
In the appendix, we give some notations about the graphical method. For detail, we refer to \cite{alesci2015hamiltonian,yang2015graphical1,yang2015graphical2,yang2017graphical} and references therein.
The 2-j symbol $\epsilon^j_{mn}$ and the 3-j symbol are represented as
\begin{equation}
\epsilon^{(j)}_{nm}=(-1)^{j+n}\delta(m,-n)=\makeSymbol{\raisebox{0.\height}{\includegraphics[width=0.2\textwidth]{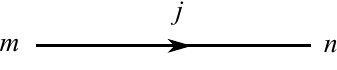}}}
\end{equation}
\begin{equation}
\left(
  \begin{matrix}
  j_1&j_2&j_3\\
  m_1&m_2&m_3
  \end{matrix}
  \right)=\makeSymbol{\raisebox{0.\height}{\includegraphics[width=0.2\textwidth]{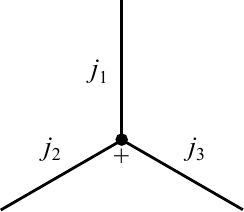}}}
\end{equation}  
For the Wigner D matrix, we define
\begin{equation}
 \makeSymbol{\raisebox{0.\height}{\includegraphics[width=0.3\textwidth]{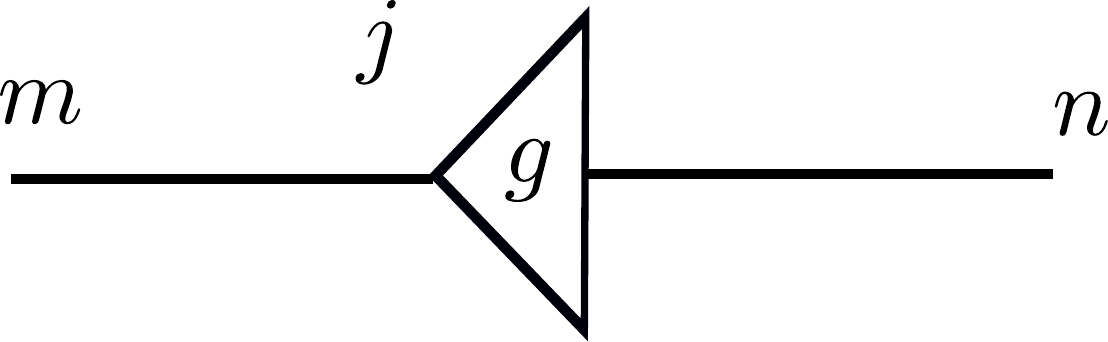}}}=:D^j(g)^m_{\ \ n}
\end{equation}

Define
$
J_\pm=\mp \frac{1}{\sqrt{2}}(J_x\pm iJ_y)
$ and $J_0:=J_z$. 
Let $|jm\rangle$ be the usual basis of the $j$-irreducible representation space of $SU(2)$. Graphically we have
 \begin{equation}
 \begin{aligned}
  \langle j m|J_\mu|j n\rangle=-W_j\left(
\begin{matrix}
j&1&j\\
n&\mu&-m
\end{matrix}  
  \right)
  =-W_j\makeSymbol{\raisebox{0.\height}{\includegraphics[width=0.2\textwidth]{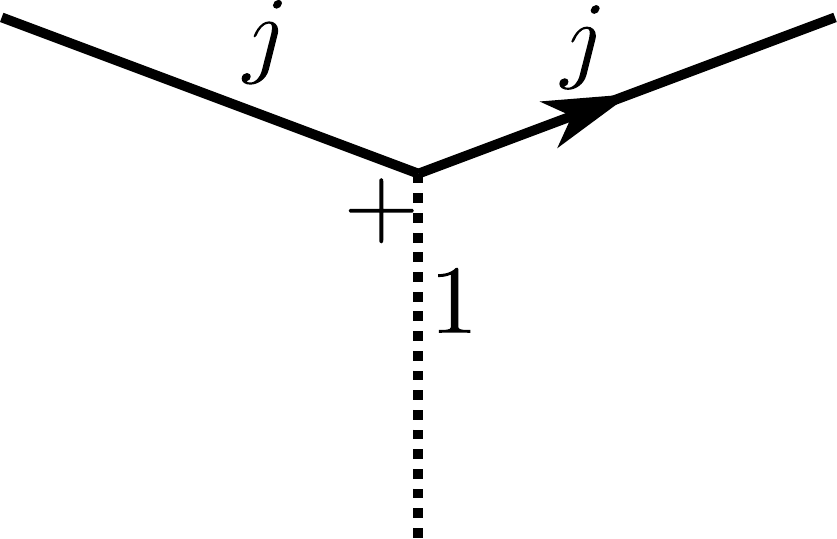}}}
  \end{aligned}
  \end{equation} 
  Let $\epsilon_{\mu\nu\sigma}$ be the totally antisymmetric matrix with $\epsilon_{-1,0,1}=1$, then
 \begin{equation}
 \begin{aligned}
 \epsilon_{\mu\nu\sigma}=\sqrt{6}\left(
\begin{matrix}
1&1&1\\
\mu&\nu&\sigma
\end{matrix} 
 \right)=\sqrt{6}\makeSymbol{\raisebox{0.\height}{\includegraphics[width=0.2\textwidth]{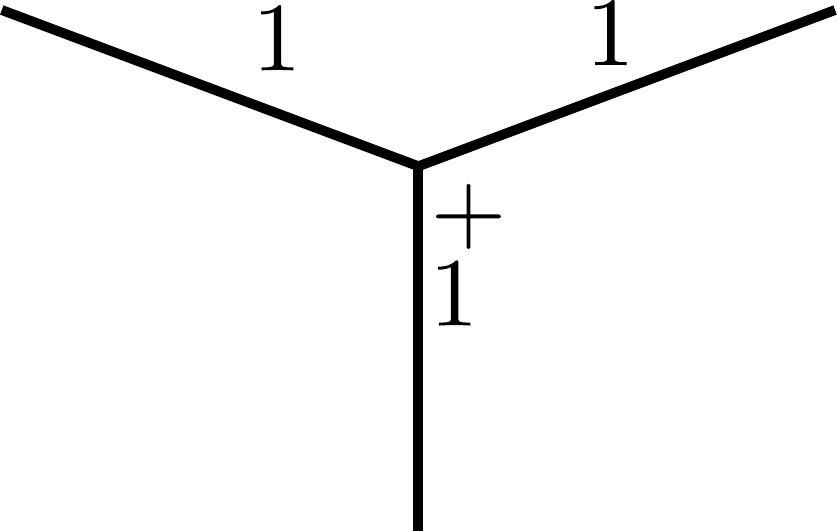}}}
 \end{aligned}
 \end{equation}

By the formula
\begin{equation}
D^{j_1}(g)^{m_1}_{\ \ n_1}D^{j_2}(g)^{m_2}_{\ \ n_2}=\sum_{J=|j-1-j_2|}^{j_1+j_2}d_J(-1)^{M-N}\left(
\begin{matrix}
j_1&j_2&J\\
m_1&m_2&-M
\end{matrix}
\right)
\left(
\begin{matrix}
j_1&j_2&J\\
n_1&n_2&-N
\end{matrix}
\right)D^J(g)^M_{\ \ N}
\end{equation}
we have
\begin{equation}
\makeSymbol{\raisebox{0.\height}{\includegraphics[width=0.2\textwidth]{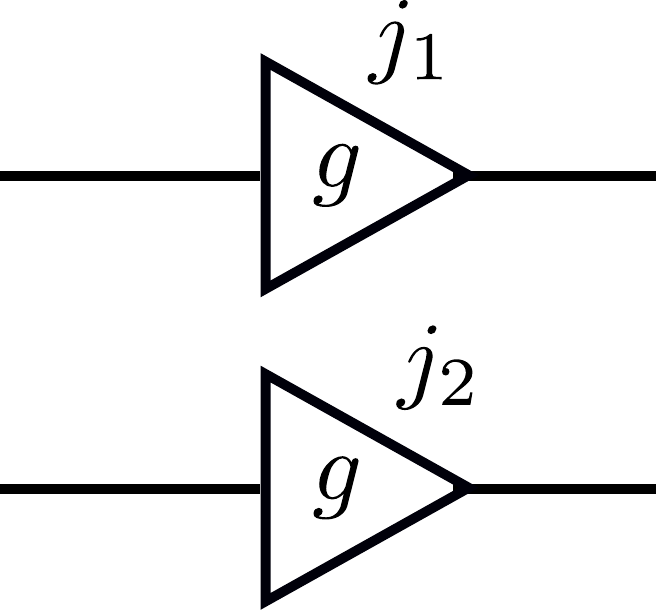}}}=\sum_{J}d_J\makeSymbol{\raisebox{0.\height}{\includegraphics[width=0.3\textwidth]{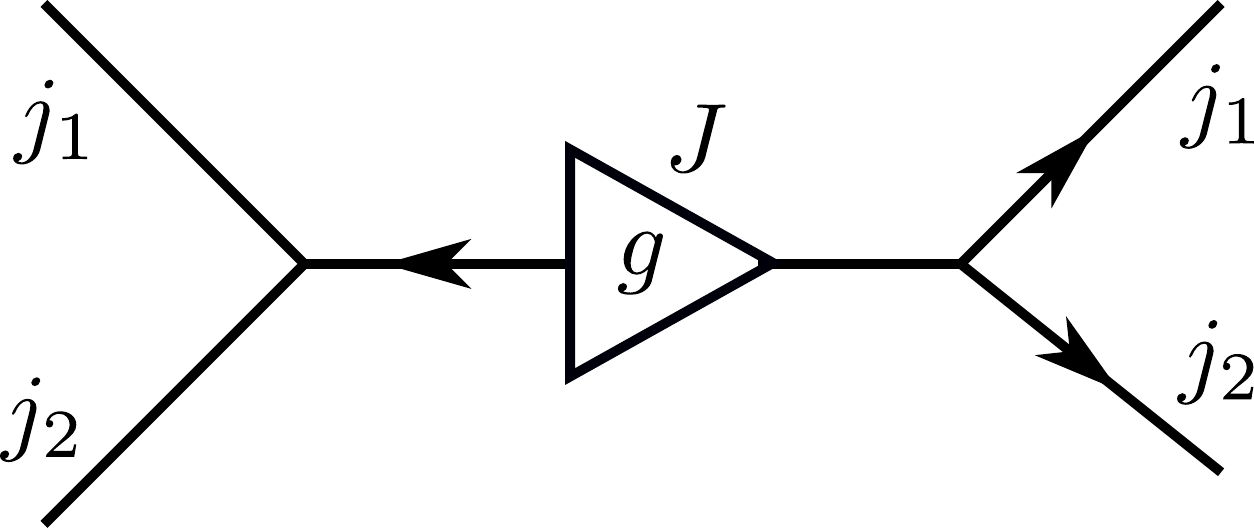}}}
\end{equation}
For the operator $J_{v,e}$ in \eqref{J}, if $t(e)=v$, we have
\begin{equation}
\begin{aligned}
J^{v,e}_\mu D^j_{mn}(g)=&J^{(R)}_\mu D^j_{mn}(g)=-\sum_{k=-j}^j\langle jm|J_\mu|jk\rangle D^j_{kn}(g) =W_j \makeSymbol{\raisebox{0.\height}{\includegraphics[width=0.2\textwidth]{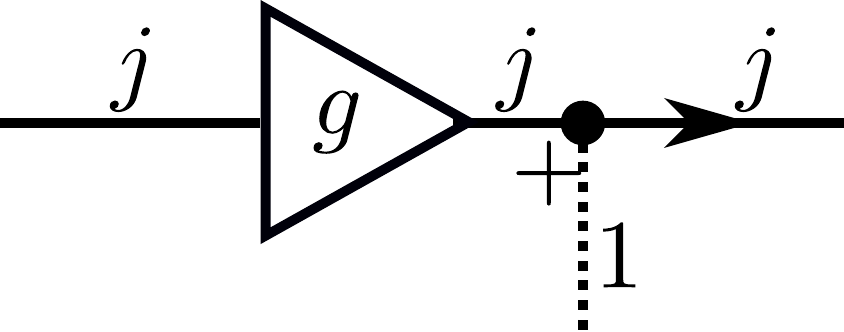}}}
\end{aligned}
\end{equation}
If $s(e)=v$, we get
\begin{equation}
\begin{aligned}
J^{v,e}_\mu D^j_{mn}(g)=&J^{(L)}_\mu D^j_{mn}(g)=D^j_{mk}(g) \langle jk|J_\mu|jn\rangle =-W_j \makeSymbol{\raisebox{0.\height}{\includegraphics[width=0.2\textwidth]{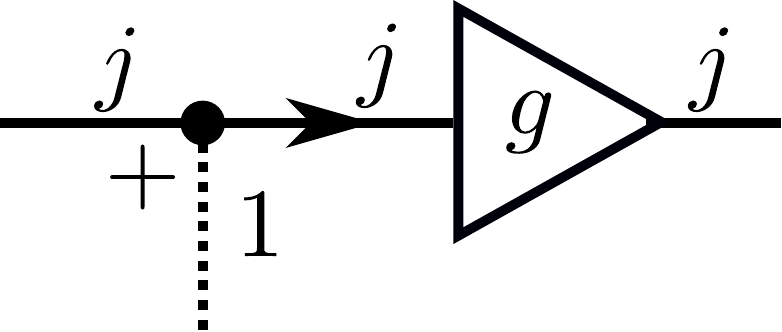}}}
\end{aligned}
\end{equation}

\section{Detail calculation of ${}^{\rm kin}H^E_v|n,\vec{j},\vec{l}\rangle$}\label{app:calculation}
We show the detail calculation of ${}^{\rm kin}H^E_v|n,\vec{j},\vec{l}\rangle$ by the graphical method introduced above. It reads
\begin{equation}
\begin{aligned}
&{}^{\rm kin}\hH^E_v~\makeSymbol{\raisebox{0.\height}{\includegraphics[width=0.3\textwidth]{first}}}
=-\kappa_1(-1)^{2l_{n+1}}3\sqrt{6}\frac{W_{j_{n+1}}^2}{W_{l_{n+1}}} \makeSymbol{\raisebox{0.\height}{\includegraphics[width=0.3\textwidth]{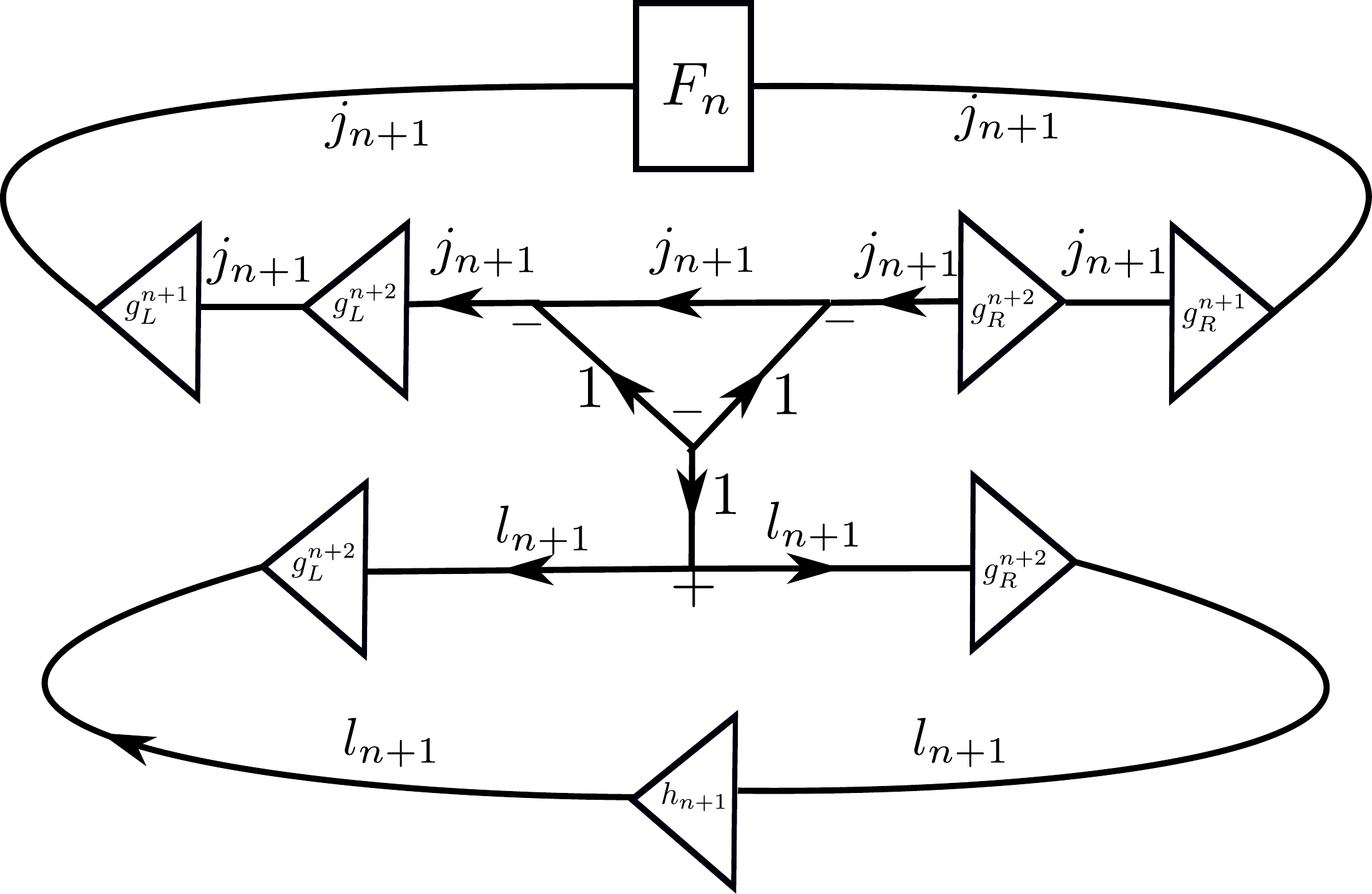}}}\\
=&-\kappa_1(-1)^{2l_{n+1}+2j_{n+2}}3\sqrt{6}\frac{W_{j_{n+1}}^2}{W_{l_{n+1}}} \makeSymbol{\raisebox{0.\height}{\includegraphics[width=0.2\textwidth]{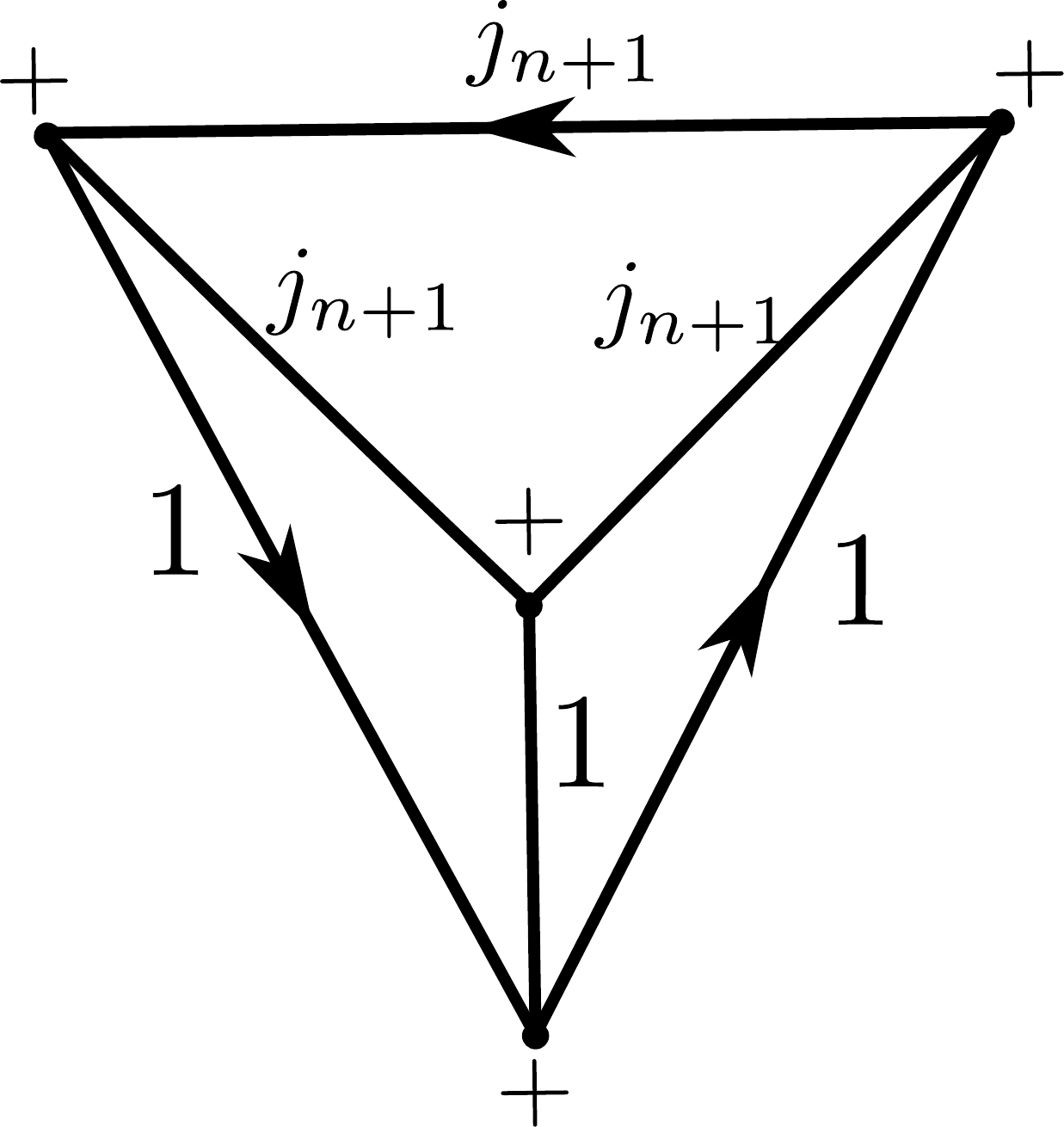}}}\sum_{j_{n+2}=|j_{n+1}-l_{n+1}|}^{j_{n+1}+l_{n+1}} d_{j_{n+2}}\times\\
&\qquad\qquad\qquad\times \makeSymbol{\raisebox{0.\height}{\includegraphics[width=0.2\textwidth]{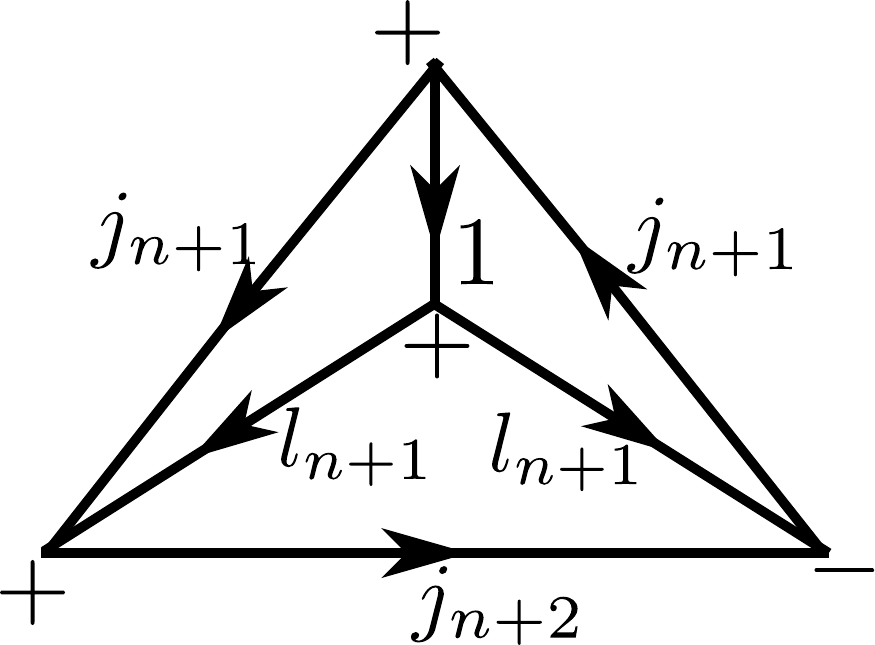}}}\makeSymbol{\raisebox{0.\height}{\includegraphics[width=0.3\textwidth]{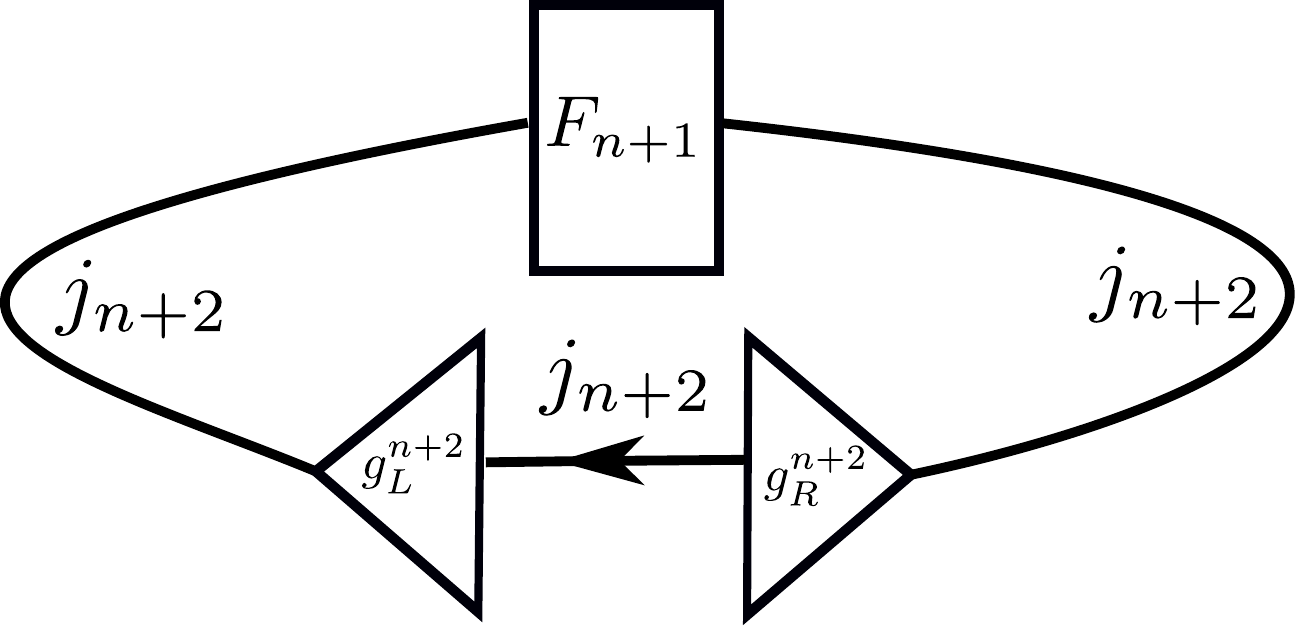}}}.
\end{aligned}
\end{equation}

By definition, we have
\begin{equation*}
\makeSymbol{\raisebox{0.\height}{\includegraphics[width=0.2\textwidth]{6j}}}=\left(
\begin{matrix}
j_{n+1}&j_{n+1}&1\\
1&1&j_{n+1}
\end{matrix}
\right)=(-1)^{2j_{n+1}}\frac{1}{\sqrt{6j_{n+1}(j_{n+1}+1)(2j_{n+1}+1)}}=\frac{(-1)^{2j_{n+1}}}{\sqrt{6}W_{j_{n+1}}}.
\end{equation*}
\begin{equation*}
\begin{aligned}
\makeSymbol{\raisebox{0.\height}{\includegraphics[width=0.2\textwidth]{6j2}}}=&(-1)^{j_{n+1}+l_{n+1}+j_{n+2}}\left(
\begin{matrix}
l_{n+1}&l_{n+1}&1\\
j_{n+1}&j_{n+1}&j_{n+2}
\end{matrix}
\right)\\
=&-\frac{\left[j_{n+1}(j_{n+1}+1)+l_{n+1}(l_{n+1}+1)-j_{n+2}(j_{n+2}+1)\right]}{2W_{j_{n+1}}W_{l_{n+1}}}\\
=:&\frac{\vec{J}_{n+1}\cdot\vec{L}_{n+1}}{W_{j_{n+1}}W_{l_{n+1}}}.
\end{aligned}
\end{equation*}

Finally, because $j_{n+1}+l_{n+1}+j_{n+2}\in \mathbb{N}$, we get
\begin{equation}
\begin{aligned}
&{}^{\rm kin}\hH^E_v~\makeSymbol{\raisebox{0.\height}{\includegraphics[width=0.2\textwidth]{first}}}=\sum_{j_{n+2}} -3\kappa_1\frac{d_{j_{n+2}}}{W^2_{l_{n+1}}}\left(\vec{J}_{n+1}\cdot\vec{L}_{n+1}\right)\makeSymbol{\raisebox{0.\height}{\includegraphics[width=0.3\textwidth]{seven}}}.
\end{aligned}
\end{equation}

Because of
\begin{equation}
\begin{aligned}
\langle \makeSymbol{\raisebox{0.\height}{\includegraphics[width=0.2\textwidth]{first}}} |\makeSymbol{\raisebox{0.\height}{\includegraphics[width=0.2\textwidth]{first}}}\rangle=\frac{1}{(d_{j_1}d_{j_2}\cdots d_{j_n})^2d_{j_{n+1}}d_{l_1}\cdots d_{l_n}},
\end{aligned}
\end{equation}
we get
\begin{equation}
\begin{aligned}
&{}^{\rm kin}H^E_v|\gamma_n,\vec{j},\vec{l}\rangle\\
=& \frac{-3\kappa_1}{l_{n+1}(l_{n+1}+1)(2l_{n+1}+1)} \sum_{j_{n+2}} \frac{\sqrt{2j_{n+2}+1}}
{\sqrt{(2j_{n+1}+1)(2l_{n+1}+1)}}\left(\vec{J}_{n+1}\cdot\vec{L}_{n+1}\right)\left|\gamma_{n+1},(\vec{j},j_{n+2}),(\vec{l},l_{n+1})\right\rangle.
\end{aligned}
\end{equation}
\section{The underlying theorem}\label{sec:C}
The underlying theorem can be found in \cite{faris1974commutators,reed1975ii} for more details. 
\begin{theorem}\label{theom:commutator-SA}
Let $\n$ be a self-adjoint operator with $\n\geq 1$. Let $\hat{A}$ be a symmetric operator with domain $D$ which is a core for $\n$. Suppose that: 
\begin{itemize}
\item[(i)] For some $c$ and all $\psi\in D$,  one has
\begin{equation}
||\hat{A}\psi||\leq c||\n\psi||.
\end{equation}
\item[(ii)] For some $d$ and all $\psi\in D$, one has
\begin{equation}\label{eq:boundedcomuter}
|(\hat{A}\psi, \n\psi)-(\n\psi, \hat{A}\psi)|\leq d||\n^{1/2}\psi||^2.
\end{equation}
\end{itemize}
Then  $\hat{A}$ is essential self-adjoint on $D$ and its closure is essentially self-adjoint on any core for $\n$.
\end{theorem} 
 
 In our case, $\h$ is the operator $\hat{A}$ in the theorem and $D=\f$ \eqref{eq:F}. By definition of $\n$ in the present work, $[\h,\n]$ is well defined on $\f$. Then \eqref{eq:boundedcomuter} can be rewritten as
\begin{equation}
|(\psi,[\h,\n]\psi)|\leq d ||\n^{1/2}\psi||^2.
\end{equation}


\end{document}